\documentclass[reqno]{amsart}
\usepackage[english]{babel}
\usepackage{amsmath}
\usepackage{amssymb}
\usepackage{enumerate}

\usepackage{amsfonts}
\usepackage{graphicx}
\newtheorem{lemma}     {Lemma}[section]
\newtheorem{thm}   [lemma]{Theorem}
\newtheorem{teorema1}   [lemma]{Theorem}
\newtheorem{prop}       [lemma]{Proposition}
\newtheorem{coro}       [lemma]{Corollary}
\newtheorem{cong1}      [lemma]{Conjecture}
\newtheorem{remark1}    [lemma]{Remark}

\numberwithin{equation}{section}

     \newcommand{\nn}{\nonumber}

\newcommand{\dis}{\displaystyle}
\def\mintone{\Xint-}

\newcommand{\mmmintone}[1]{{\dis{\int\kern -.38cm
-}}_{\kern-.21cm\substack{#1}}\;\;}
\newcommand{\mmmintwo}[2]{{\dis{\int\kern -.43cm
-}}_{\kern-.21cm\substack{#1}}^{\substack{#2}}\;\;}
\newcommand{\submint}{{\scriptstyle{\int\kern -.66em -}}}
\newcommand{\submintone}[1]{{\scriptstyle{\int\kern -.66em
-}}_{\scriptscriptstyle{\kern-.21em\substack{#1}}}}
\newcommand{\fracmint}{{\textstyle{\int\kern -.88em -}}}
\newcommand{\fracmintone}[1]{{\textstyle{\int\kern -.88em
-}}_{\scriptscriptstyle{\kern-.21em\substack{#1}}}\;}

\def\mintone{\protect\mmmintone}

\newcommand{\eps}{\epsilon}

\newcommand{\ga}{\gamma}

\newcommand{\Om}{\Omega}

\newcommand{\si}{\sigma}

\newcommand{\la}{\lambda}
\newcommand{\La}{\Lambda}

\newcommand{\nada}[1]{}

\begin{document}



\title[Fourier law and phase transitions]
{Fourier law, phase transitions and the stationary Stefan problem}

\author{A. De Masi}
\address{Anna De Masi,
Dipartimento di Matematica, Universit\`a di L'Aquila \newline
\indent L'Aquila, Italy}
\email{demasi@univaq.it}

\author{E. Presutti}
\address{Errico Presutti,
Dipartimento di Matematica, Universit\`a di Roma Tor Vergata \newline
\indent Roma, 00133, Italy}
\email{Presutti@mat.uniroma2.it}

\author{D. Tsagkarogiannis}
\address{Dimitrios Tsagkarogiannis,
Dipartimento di Matematica, Universit\`a di Roma Tor Vergata \newline
\indent Roma, 00133, Italy}
\email{tsagkaro@mat.uniroma2.it}

\maketitle

\begin{abstract}
We study the one-dimensional stationary
solutions of an integro-differential equation
derived by Giacomin and Lebowitz from Kawasaki dynamics in
Ising systems with Kac potentials,  \cite{GiacominLebowitz}.
We construct stationary solutions with non zero current
and prove
the validity of the Fourier law in the thermodynamic limit showing that
below the critical temperature the limit
equilibrium profile has a discontinuity (which defines the position of the interface)
and  satisfies a stationary free boundary Stefan
problem.
Under-cooling and over-heating effects are also studied.  We show that
if metastable values are imposed at the boundaries
then the mesoscopic stationary profile
is no longer monotone and therefore
the Fourier law is not satisfied. It regains however
its validity in
the thermodynamic limit where the limit profile is again monotone away from the interface.
\end{abstract}

\section{\hskip.2cm Introduction}
\label{intro}

When hydrodynamic or thermodynamic limits are performed
in systems which are in the
phase transitions regime
we may observe perfectly
smooth profiles develop singularities  with the appearance of sharp interfaces.
We shall study the phenomenon in
stationary non equilibrium states which carry non zero steady
currents, the general context is the one where the Fourier law applies, but here  it is
complemented by
a free boundary problem due to the presence of  interfaces.
We work at the mesoscopic level considering a model
which has been derived in  \cite{GiacominLebowitz} from
Ising systems
with Kac potentials and Kawasaki dynamics, and derive in the hydrodynamic limit
macroscopic profiles
with an interface
which satisfy a stationary Stefan problem and obey the Fourier law.

The mesoscopic model is defined in terms of
a  free energy
functional, the  Lebowitz and Penrose [L-P] functional (see \eqref{e2.1} in the next section)
which is a non local version of the scalar Ginzburg-Landau (or Allen-Cahn
or Cahn-Hilliard) functional.
Its  thermodynamic free energy
density
is obtained by minimizing the  L-P
functional over profiles with fixed total magnetization density
and then taking the thermodynamic limit where the spatial size of the
system diverges. It is found that the phase
diagram (of
free energy
density versus magnetization density) obtained in this way
has a non trivial
flat interval $[-m_\beta,m_\beta]$ (indicative of a phase transition)
when the inverse temperature $\beta$ is above the
critical value (equal to 1 here). This is in qualitative and quantitative agreement
with the thermodynamics of the underlying
Ising model with Kac potentials,
see Chapter 9 in \cite{presutti} and references therein.
The axiomatic theory for such phase diagrams  predicts that the values
inside $(-m_\beta,m_\beta)$ do not  appear in any stationary local equilibrium state, so that
a macroscopic magnetization density profile
will have a discontinuity  if it  assumes
values both smaller than $-m_\beta$ and
larger than $m_\beta$.

This is just what we see. We fix  $\beta>1$ and study the stationary
solutions of the equations of motion  $\dis{\frac{dm}{dt}=-{\rm div}\;I}$, i.e.\
${\rm div}\;I=0$,
$I$ the local current (of the conserved order parameter, the magnetization density $m$ here).
By a gradient flow assumption on its constitutive law,
$I$ is supposed proportional to the gradient
of the functional derivative of
the  L-P functional:  due to the non local structure
of the latter, $\dis{\frac{dm}{dt}=-{\rm div}\;I}$ is
an integro-differential
equation  (see \eqref{e2.18} in the next section), which is the same as the one derived by
Giacomin and Lebowitz from the Ising system,  \cite{GiacominLebowitz}, and which
has been much studied in the
past years,  \cite{LOP}, \cite{asselah}, \cite{GLM}.
We look for solutions of  ${\rm div}\; I=0$ with a planar symmetry thus reducing to
a one dimensional problem and
prove existence
and smoothness of solutions with a steady non-zero current.
However in the hydrodynamic limit
where the size $L$ of the system diverges,
the stationary profile, once expressed
in macroscopic space units (i.e.\ proportional to $L$),
is proved to converge to a  discontinuous limit profile,
solution  of a stationary free boundary problem, the stationary
Stefan problem,
in agreement with the axiomatic macroscopic theory.  The mesoscopic theory
is in this respect in complete agreement with the macroscopic one,
the mesoscopic profiles are smooth versions of the  macroscopic ones,
they are  monotone as well and  the
current is proportional to [minus] the magnetization density gradient
in agreement with the Fourier law which we may then say to be valid at the
mesoscopic level as well.

The mesoscopic theory has  however a richer and  more complex structure
even in the macroscopic limit.  This is seen for instance if we impose boundary conditions
which force  metastable values at the boundaries, the metastable region
being made of two separate intervals called the plus and the minus metastable phases
(according to the sign of the magnetization) which (together with the spinodal region) are
 contained in the ``forbidden region''
$(-m_\beta,m_\beta)$. With  boundary conditions one in the minus,
the other in the plus metastable phases the mesoscopic stationary
magnetization density profiles are not monotone anymore. We have the
``paradoxical'' result of a positive [magnetization] current when
also the total magnetization gradient is positive having fixed at
the left and right respectively a negative and a positive metastable
value of the magnetization. The mesoscopic stationary profile  is
then first decreasing, then increasing and then again decreasing.
The Fourier's law is therefore not satisfied but, in the
thermodynamic limit, the region where the profile increases shrinks
to a point, which is where the limit profile has a discontinuity (a
sharp interface). Elsewhere the profile is always decreasing in
agreement with the Fourier's law (as the current is positive). The
stationary profile has therefore values all in the metastable region
(except at the interface which macroscopically is only a point). All
the  issues presented in this introduction are discussed in some
more details in the next section, proofs are given in the remaining
ones.

\vskip2cm

\section{\hskip.2cm Model, backgrounds  and main results}
\label{sec:e2}

The free energy functional to which  we have been referring so far is
defined on functions $m\in L^{\infty}(\La,[-1,1])$, $\La$
a bounded measurable subset of $\mathbb R^d$, as
   \begin{eqnarray}
&& F_{\beta,\La}(m|m_{\La^c})=F_{\beta,\La}(m)+\frac 12\int_{\La}\int_{\La^c} J(x,y)
[m(x)-m_{\La^c}(y)]^2\nn\\
&& F_{\beta,\La}(m)=\int_{\La} \phi_\beta(m)
+\frac 14\int_{\La}\int_{\La} J(x,y)
[m(x)-m(y)]^2
    \label{e2.1}
    \end{eqnarray}
where $J(x,y)=J(|x-y|)$ is a smooth, translational
invariant,
probability kernel of range 1; $m_{\La^c}\in
L^{\infty}(\La^c,[-1,1])$ is a fixed external profile and
   \begin{equation}
    \label{e2.2}
\phi_\beta(m) = -\frac 12 m^2-\frac 1\beta
S(m),\;\;
-S(m)=\frac {1+m} 2\log\Big(\frac{1+m} 2\Big)+ \frac {1-m}
2\log\Big(\frac{1-m} 2\Big)
    \end{equation}
To simplify the analysis we suppose $\La$ a cube and consider Neumann boundary conditions, namely the functional
   \begin{eqnarray}
F_{\beta,\La}^{\rm neum}(m)=\int_{\La} \phi_\beta(m)
+\frac 14\int_{\La}\int_{\La} J^{\rm neum}(x,y)
[m(x)-m(y)]^2
    \label{e2.1.1}
    \end{eqnarray}
where $\dis{J^{\rm neum}(x,y) = \sum_{z\in R_\La(y)} J(x,z)}$ with $R_\La(y)$
the set image of
$y$ under reflections of the cube $\La$ around its faces.
In $d=1$, if $\La=\eps^{-1}[-1,1]$,
$J^{\rm neum}(x,y)=J(x,y) + J(x,2\eps^{-1} -y) +  J(x,-2\eps^{-1} - y)$ ($\eps>0$ is a scaling parameter which will
vanish in the thermodynamic limit).  With minor modification
what follows in the next item ``Equilibrium thermodynamics'' holds
as well for general boundary conditions as
those considered  in \eqref{e2.1}.

    \vskip1cm

\centerline {\em Equilibrium thermodynamics of the mesoscopic model.}
 \nopagebreak

 \vskip.1cm
 \noindent
(The statements in this
paragraph are proved in Section 6.1 of \cite{presutti}).
The thermodynamic free energy density $a_\beta(s)$, $s\in [-1,1]$, is defined as
    \begin{equation}
    \label{e2.3}
a_\beta(s):= \lim_{\La \to \mathbb R^d}
\inf\Big\{ F^{\rm neum}_{\beta,\La}(m)\; \Big|\; \mintone{\La} m =  s\Big\}
    \end{equation}
The limit on the r.h.s.\ indeed exists and it is equal to:
    \begin{equation}
    \label{e2.4}
a_\beta= \phi^*_{\beta}=\;\text{convex envelope of $\phi_\beta(\cdot)$}
    \end{equation}
$\phi^*_\beta\equiv \phi_\beta$ when $\beta\le 1$ and
$\phi^*_\beta\ne \phi_\beta$ when $\beta> 1$.  More precisely
let $m_\beta$ be
the positive solution of
    \begin{equation}
    \label{e2.5}
m_\beta= \tanh \{\beta m_\beta\}, \;\;  \beta>1
    \end{equation}
then   $\phi^*_\beta(s)$, $s\in (-m_\beta,m_\beta)$, is  constant and strictly smaller
than $\phi_\beta(s)$,  while
$\phi^*_\beta(s)=\phi_\beta(s)$ elsewhere.  The values of the magnetization in the interval
$ (-m_\beta,m_\beta)$ are ``forbidden''.  This is best seen working
in
the grand canonical ensemble (in other words, using Lagrange multipliers).
To this end we add a constant magnetic field $h$ so that  the
free energy functional becomes
   \begin{equation}
    \label{e2.6}
F^{\rm neum}_{\beta,h,\La}(m)=F^{\rm neum}_{\beta,\La}(m) -h \int_{\La}  m
    \end{equation}
The grand canonical thermodynamic pressure
$p_\beta(h)$ is defined by a minimization problem  without constraints:
    \begin{equation}
    \label{e2.7}
p_\beta(h)=   \lim_{\La \to \mathbb R^d} \sup\Big\{- F^{\rm neum}_{\beta,h,\La}(m) \;\Big|\;  m \in L^\infty(\La,[-1,1])\Big\}
    \end{equation}
Existence of the limit is again a fact and the two thermodynamics
defined by the free energy $a_\beta$ and by the pressure $p_\beta$
are equivalent, a property called in statistical mechanics
``equivalence of ensembles''. Namely $p_\beta$ and $a_\beta$ are
inter-related as in thermodynamics being one the  Legendre transform
of the  other:
    \begin{equation}
    \label{e2.8}
p_\beta(h)=  \sup\big\{hs -a_\beta(s) \;\big|\; s\in [-1,1]\big\},\quad
a_\beta(s)=  \sup\big\{hs - p_\beta(h)\;\big|\; h\in \mathbb R\big\}
    \end{equation}
For any $\beta>1$ and any $h\in \mathbb R$ any maximizer of \eqref{e2.7} at least for  $\La$
large enough is a
constant  function equal to $m_{\beta,h}$ where  $m_{\beta,h}$ is the solution of
the mean field equation
   \begin{equation}
    \label{e2.9}
m_{\beta,h}= \tanh \{\beta (m_{\beta,h}+h)\}
    \end{equation}
which minimizes $\phi_\beta(s)-hs$ and therefore it is
not in $(-m_\beta,m_\beta)$,
the values in  $(-m_\beta,m_\beta)$ ``are therefore forbidden''.

    \vskip1cm

\centerline {\em Gibbsian equilibrium thermodynamics.}
 \nopagebreak
 \vskip.1cm
 \noindent
The thermodynamics obtained above is in qualitative and quantitative agreement with
the thermodynamics of the underlying microscopic model, i.e.\ the Ising  system
with Kac potential.  The Gibbs canonical
equilibrium free energy $f_{\beta,\ga}(m)$ is defined as
   \begin{equation}
    \label{e2.9.1}
f_{\beta,\ga}(m): =  \lim_{\delta\to 0}\lim_{\La_n \to \mathbb Z^d} \frac{-1}{\beta|\La_n|}\log Z_{\La_n,\beta,\ga}
    \end{equation}
    \begin{equation*}
    Z_{\La_n,\beta,\ga}= \sum_{\si_{\La_n}\in \{-1,1\}^{\La_n}}
\text{\bf 1}\Big(|\sum_{x\in \La_n} (\si_{\La_n}(x) - m)| \le \delta|\La_n|\Big) e^{- \beta H_{\ga,\La_n}(\si_{\La_n})}
\end{equation*}
where $\La_n$ is  a sequence of increasing cubes and
   \begin{equation}
    \label{e2.9.2}
 H_{\ga,\La}(\si_\La) = -\frac 12\sum_{x\ne y \in \La} J_\ga(x,y)\si_\La(x)\si_\La(y),\quad
 J_\ga(x,y)= \ga^d J_\ga(\ga|x-y|)
    \end{equation}
(Same free energy is obtained for more general regions and boundary conditions).
As discussed in Chapter 9 of \cite{presutti} in $d\ge 2$ for any $\beta>1$  and $\ga>0$ small enough,
$f_{\beta,\ga}(m)$  is  flat in an interval $[-m_{\beta,\ga},m_{\beta,\ga}]$ and
$m_{\beta,\ga} \to m_\beta$ as $\ga\to 0$.  The original result has been proved in \cite{CP}
and \cite{BZ} while the fact that in any $d\ge 1$, $\dis{\lim_{\ga\to 0}f_{\beta,\ga}(m)
=a_\beta(m)}$ is much older and proved by Lebowitz and Penrose, \cite{LP}.

    \vskip1cm

\centerline {\em  Axiomatic non equilibrium macroscopic theory.}
 \nopagebreak
 \vskip.1cm

\noindent The basic postulates are (i)--(iv).

\noindent
(i)\; {\it local equilibrium and
barometric formula}. The free energy of a macroscopic
profile $m$ in the macroscopic (bounded) region $\Om\subset \mathbb R^d$ is given by the local functional:
   \begin{equation}
    \label{e2.10}
F^{\rm macro}_{\beta, \Om}(m) := \int_\Om a_\beta(m),\quad m \in L^\infty(\Om,[-1,1])
    \end{equation}
(ii)\; {\it gradient dynamics}. The  evolution equation in the interior of $\Om$ is the conservation law  $(D$ below denoting functional derivative)
   \begin{equation}
    \label{e2.11}
\frac{dm}{dt}= -\nabla j,\quad j=-\chi \nabla D F^{\rm macro}_{\beta, \Om} = -\chi \nabla a'_\beta,\; a'_\beta(s):=
\frac{da_\beta(s)}{ds}
    \end{equation}
(iii) \; {\it mobility coefficient}. $\chi$ is a mobility coefficient which depends on the
dynamical characteristics  of the system, we take
   \begin{equation}
    \label{e2.12}
\chi(s)= \beta(1-s^2)
    \end{equation}
as this is what found
when deriving \eqref{e2.11} from the Ising spins, \cite{GiacominLebowitz}-\cite{LOP}.

The usual setup for Fourier law has  $\Om$
a parallelepiped with different
values of the order parameter imposed on its right and left faces and Neumann
(or periodic) conditions
on the other ones.
By the  planar symmetry
the problem becomes one dimensional and from now on we shall restrict
to $d=1$ taking $\Om=[-\ell,\ell]$.    The stationary profiles $m(x)$, $x\in (-\ell,\ell)$, verify
   \begin{equation}
    \label{e2.13}
D_\beta \frac{dm}{dx}= -j=\text{\rm constant},\; D_\beta(m)= \chi
(m) a''_\beta(m)
    \end{equation}
and are determined  for instance by  Dirichlet boundary conditions
at $\pm \ell$, namely $m(x)\to m_{\pm}$ as $x\to \pm \ell$.  To have
an increasing profile we shall suppose that $-1<m_-<-m_\beta$ and
$1>m_+>m_\beta$, the opposite case being recovered by symmetry. When
$\beta<1$ the above is well posed as $a_\beta''>0$   but if
$\beta\ge 1$ the denominator vanishes.  The macroscopic theory then
needs a further postulate:

\noindent (iv) \; {\it The stationary
Stefan problem.}  There are $x_0\in (-\ell,\ell)$ and $j<0$ so that there is a solution $m(x)$ of \eqref{e2.13} in $(-\ell,x_0)$
with boundary values $m_-$ and $-m_\beta$ and in $(x_0,\ell)$
with boundary values $m_\beta$ and $m_+$.   The current
$\dis{-\chi(m(x))\frac{d}{dx}
a'_\beta(m(x))}$ being equal to $j$ is constant through the interface $x_0$: conservation
of mass would otherwise impose a motion of the interface against the assumption
that the profile is stationary.  Observe also that since $a'_\beta(m)$ is an increasing function of $m$ in $[-m_\beta,m_\beta]^c$
and since $h(x)$ is increasing (if $j<0$) then $m(x)$ is also increasing.

\vskip.2cm

A different formulation of the problem is however more
convenient for our purposes. We start by a change
of variables, going from $m$ to $h$.  There is a one to one correspondence between
$m$ and $h$ when
$\{m\ge m_\beta\}$ and $\{h\ge 0\}$ and also  when $\{m\le -m_\beta\}$ and $\{h\le 0\}$.
The correspondence
is  given
in one direction by \eqref{e2.9},
and in the other by  $h=a'_\beta(m)$.  Expressed in terms of the magnetic field, \eqref{e2.13}
becomes
   \begin{equation}
    \label{e2.14}
 h(x)= \int_{x_0}^x  \frac{-j }{ \chi(m)},\quad  m= (a'_\beta)^{-1}(h)
     \end{equation}
namely $m$ is regarded as a function of $h$ obtained by inverting
$h=a'_\beta(m)$ and $\chi(m)=\chi(m(h))$ becomes   a function of $h$
as well.
\eqref{e2.14} is then an integral equation in $h(\cdot)$
where  however  $x_0$ and $j$ are also unknown: they must be
determined by imposing the boundary conditions $h(\pm
\ell)=h_{\pm}:=a'_\beta(m_{\pm})$. All this suggests a new
formulation (alternative to the Dirichlet problem) where we assign
$x_0$ and $j$ instead of $m_{\pm}$. In this way the Stefan problem
is written in a compact way as in \eqref{e2.14} above which is now a
``pure'' integral equation for $h(\cdot)$ with $x_0$ and $j$ known
data. We shall mostly use in the sequel this latter formulation when
proving that the Stefan problem with assigned $x_0$ and $j$ can be
derived from the mesoscopic theory.

As a difference with the Dirichlet problem, in the ``$x_0,j$ problem''  there is
no  ``global existence theorem'', in the sense that
given   $x_0$ and $j$ there are no solutions if $\ell$ is too large.  Indeed
\eqref{e2.14} with $x_0=0$ and $j< 0$ has a ``maximal solution''
$(h_j(x),m_j(x))$. Namely there is a bounded interval $(-\ell_j,\ell_j)$ such that
   \begin{equation}
    \label{e2.23}
\lim_{x\to \pm \ell_j}m_j(x)=\pm 1,\quad \lim_{x\to \pm \ell_j}h_j(x)= \pm \infty
    \end{equation}
\eqref{e2.14}  has no solution if $\ell>\ell_j$ while any other solution of
\eqref{e2.14} with the same $j$ is obtained, modulo translations,
by  restricting
the maximal solution to a  suitable interval contained in  $(-\ell_j,\ell_j)$.
The value $\ell_j$ is strictly finite because
the solution $m(h)$ of $m=\tanh\{\beta h +\beta m\}$ when $h\to \infty$ and
$m(h)\to 1$ is to first order given by $\dis{\frac{dm}{dh} \approx \beta(1-m^2)}$.
Thus  $\dis{\frac{dm}{dx} \approx -j}$ in
\eqref{e2.13} when $m\approx 1$ hence $m(\cdot)$ converges to 1
linearly with slope $-j$ (recall $j<0$).  The collection of all the maximal solutions
$(h_j(x),m_j(x))$ when $j\in \mathbb R\setminus\{0\}$  determines in the sense
explained above  all the possible solutions of
\eqref{e2.14}. Since $\ell_j\to 0$ as $j\to \infty$ and $\ell_j\to \infty$ as $j\to 0$ it then follows
that for any $\ell$ the Dirichlet problem with data $m_{\pm}$ at $\pm \ell$
($m_+\ne m_-$, $m_{\pm}$ in the complement of $[-m_\beta,m_\beta]$)
can
be obtained as described above from the collection of all the maximal solutions. By taking limits  we
can also include $m_\beta$ and $-m_\beta$.

\noindent By restricting to intervals strictly contained in the maximal interval $[-\ell_j,\ell_j]$ the solution
$(h,m)$ of \eqref{e2.14} is  smooth, $\|m\|<1$, $\chi(m)$ bounded away from 0 and $\|h\|<\infty$.  These are the properties of the macroscopic solution which
will be repeatedly used in the sequel.

\vskip1cm

\centerline{\em Stationary mesoscopic profiles.}
 \nopagebreak
 \vskip.1cm

\noindent  Dynamics is defined using the same
postulate of the macroscopic theory, namely it is
the gradient flow of the free energy functional which, in the
mesoscopic theory is \eqref{e2.1.1} (supposing again Neumann conditions).
The gradient flow is $(D$ below denoting functional derivative)
   \begin{eqnarray}
    \label{e2.18}
&&\frac{dm}{dt}= -\nabla I,\quad  I=-\chi \nabla \Big(D F_{\beta, \La} \Big)
 \\&& I=
-\chi \nabla \Big(\frac1{2\beta}\log \frac{1+m}{1-m}
-\int J^{\rm neum}(x,y) m(y)\,dy\Big)
        \nn
    \end{eqnarray}
With the choice $\chi=\beta(1-m^2)$ (that we adopt hereafter) \eqref{e2.18}
becomes the one found in \cite{GiacominLebowitz} from the Ising spins.  We
suppose again a planar symmetry to reduce to
 one dimension, take $\La=\eps^{-1}[-\ell,\ell]$ interpreting $\eps^{-1}$ as the ratio of
macroscopic and mesoscopic lengths so that \eqref{e2.18} becomes
   \begin{equation}
    \label{e2.19}
\frac{dm}{dt}= -\frac{d}{dx}\Big( -\frac{dm}{dx} + \beta(1-m^2)\frac{d}{dx} J^{\rm neum}*m\Big)
    \end{equation}
As in the macroscopic theory it is now convenient to change variables. Define
$h(x)$  as
   \begin{equation}
    \label{e2.20}
h:= \frac1{2\beta}\log \frac{1+m}{1-m}
-J^{\rm neum}*  m
    \end{equation}
Then the current $I$ in \eqref{e2.18} has the expression
   \begin{equation}
    \label{e2.21}
 I= - \chi(m) \frac{dh}{dx},\quad m= \tanh\{ \beta J^{\rm neum}*  m +\beta h\}
    \end{equation}
The stationary problem in the $x_0,j$ formulation is
then the following. Given any $x_0\in (-\ell,\ell)$ and $j<0$,
find  $m$ and $h$ so that
   \begin{equation}
    \label{e2.22}
m= \tanh\{ \beta J^{\rm neum}*  m +\beta h\},\quad   h(x)=
\int_{\eps^{-1} x_0}^x  \frac{-\eps j }{ \chi(m)}
    \end{equation}
We  first consider the simpler antisymmetric case  where $m$ and $h$ are both odd functions.

\vskip.5cm

             \begin{thm}
        \label{thme2.1}
Let $j\ne 0$, $x_0=0$, $\ell>0$   and smaller than $\ell_j$ (see
\eqref{e2.23}). Then for any  $\eps>0$ small enough there is an
antisymmetric pair $(h_\eps(x),m_\eps(x))$ which solves
\eqref{e2.22} in $\eps^{-1}(-\ell,\ell)$ and
$(h_\eps(\eps^{-1}x),m_\eps(\eps^{-1}x))$ converges in sup-norm as
$\eps\to 0$ to the pair $(h(x),m(x))$ solution of the Stefan problem
\eqref{e2.14}. Moreover $h_\eps$ and $m_\eps$ are both strictly
increasing if $j<0$ and strictly decreasing if $j>0$.

             \end{thm}

\vskip.5cm

\noindent {\em Remarks.}  (a)\; Theorem \ref{thme2.1} is proved in Section \ref{sec:e3}
and  in
Appendix \ref{appA}, \ref{appB} and \ref{appC}.
The proof is based on finding the fixed
point of the following map: given a function $h$ solve the first one in
\eqref{e2.22} to get  $m$ and use the second one to find the new $h$.  Existence of
a fixed point is proved by showing convergence of the iterates $h_n$ and of the corresponding $m_n$.
Since $x_0=0$ if we start with an antisymmetric function, the whole orbit remains antisymmetric
and indeed
the limit macroscopic solution is antisymmetric as well. As we shall see restricting to
the space of odd
functions greatly simplifies
the problem. We
start the iteration from
a profile $m_0$ which is almost a
fixed point: $m_0$ is in fact
the [scaled by $\eps^{-1}$] macroscopic solution away from $0$ while it is
equal to the ``instanton'' (see Section \ref{sec:e3}) in a neighborhood of $0$.
We shall  prove
that all  the profiles $m_n$ obtained by iterating \eqref{e2.22}
are  contained in a small neighborhood
of $m_0$ and that the iterates converge to a limit profile $m$; also the corresponding magnetic fields $h_n$
are proved to converge to a limit $h$  and the pair $(h,m)$ is the desired fixed point
which  solves \eqref{e2.22}.
The crucial point in the analysis is to control the change $\delta m$ of $m$  in the first equality in
\eqref{e2.22}
when we slightly vary  $h$ by $\delta h$.   To
linear order $\delta m$ and $\delta h$ are related by
$(A_{h,m}-1)\delta m= -p_{h,m}\delta h$ where $A_{h,m}=p_{h,m}J*$,
$J*$ the convolution operator with kernel $J$,
and
   \begin{eqnarray}
  \label{e2.1222}
&&p_{h,m}= \frac{\beta}{\cosh^2\{\beta J^{\rm neum}*m+\beta h\}}
\\&& p_{h,m}= \chi(m) \quad \text{if $m= \tanh\{\beta J^{\rm neum}*m+\beta h\}$ }
  \label{e4.1.1}
    \end{eqnarray}
(the equality  $p_{h,m}=\chi(m)$ in \eqref{e4.1.1} will be often exploited in
the sequel).  Thus
$\delta m = L_{h,m}^{-1} (-p_{h,m}\delta h)$ provided $L_{h,m}:=A_{h,m}-1$ is invertible.
In \cite{DOP} it is shown that
the largest eigenvalue of $L_{h,m}$
converges to 0 as $\eps\to 0$ and that there is a
spectral gap bounded away from 0 uniformly in $\eps$.  By restricting
to odd functions the leading eigenvalue disappears
and the invertibility problem can then be solved.
As clear from this outline the proof does not give uniqueness
which is left open.

  (b)\; The choice of Neumann conditions simplifies the analysis but other
conditions (provided they preserve antisymmetry)
may be treated as well unless they contrast with the macroscopic value
of the magnetization imposed by  $j$, in which case
boundary layers may appear, which are instead absent
with Neumann conditions.

   (c)\;  With Neumann conditions the non local
convolution term is completely defined, but since the evolution
involves also derivatives  other conditions are needed to determine
the solution: our choice was to fix $j$ and $x_0$.  Dirichlet
conditions would instead prescribe the limits $m_{\pm}$ of $m(x)$ as
$x\to \pm \eps^{-1}\ell$. There are here two types of boundary
conditions, those which fix $m$ outside the domain and are used to
define the convolution (in our case replaced by Neumann conditions)
and those which prescribe the values of $m$ when going to the
boundary from the interior (in our case are replaced by $j$ and
$x_0$).  The distinction is not as clear in other models as for
instance in the Cahn-Hilliard equation where more parameters are
involved, we are indebted to N. Alikakos and G. Fusco for many
enlightening discussions on such issues.

  (d)\; In this paragraph it is convenient to refer to
  Dirichlet boundary conditions. It follows immediately from
\eqref{e2.11}
that the critical points
of the functional are stationary, i.e.\ such that
the derivative vanishes, $DF_{\beta,\La}(m)=0$, $\La=\eps^{-1}[-\ell,\ell]$.
They are in fact special solutions of \eqref{e2.22}: those
with $j=0$ and hence $h=0$, thus  solutions of the mean field equation
$m=\tanh\{\beta J*(m+m_{\La^c})\}$, when $m_{\La^c}$ is
fixed outside $\La$.
In this case the
limit values of $m$ when $x\to \partial \La$ from the
interior cannot be prescribed
independently, they are generally different from those obtained
going to $\partial \La$ from the outside by using $m_{\La^c}$.  If we want
different boundary values (from the inside) than those produced
by solving $DF_{\beta,\La}(m)=0$,  we must look for solutions with a current and we are
back to the problem considered in this paper.  The solutions
of the Dirichlet problem with and without currents
are qualitatively different.  In the former case
 there is a sensitive dependence on the boundary values, even macroscopically away
from the boundaries while, when the  current is zero,
we see the familiar exponential relaxation towards the stable phases.

\vskip.5cm

We have a slightly weaker result when $x_0\ne 0$ as
we need in our proofs to replace the condition $h(\eps^{-1}x_0)=0$
by an integral one, namely $\dis{\int_{-\eps^{-1}\ell}^{\eps^{-1}\ell} h u^*=0}$,
where $u^*$ (whose dependence on $\eps$ is not made explicit) is
a suitable positive function on $\mathbb R$, symmetric
around $\eps^{-1}x_0$ and which decays exponentially as
$|x-\eps^{-1}x_0| \to \infty$ uniformly in $\eps$ (if $u^*$
were a delta we  would then be back to the condition
$h(\eps^{-1}x_0)=0$).
We do not control the exact
mesoscopic location of
the zeroes of the magnetization profile $m$ and of the magnetic
field profile $h$, however they differ from $\eps^{-1}x_0$
by  quantities which vanish faster than any power of $\eps$
as $\eps\to 0$:

\vskip.5cm

             \begin{thm}
        \label{thme2.2}
Let $j\ne 0$, $\ell \in (0,\ell_j)$ and $x_0\ne 0$ in
$(-\ell,\ell)$.  Then for any  $\eps>0$ small enough there is a
pair $(h_\eps,m_\eps)$ which solves
\eqref{e2.22} in $\eps^{-1}(-\ell,\ell)$. $h_\eps(x_\eps)=0$ where $
x_\eps\in \eps^{-1}(-\ell,\ell)$ and   $\eps x_\eps \to x_0$ (see
\eqref{eG.26} in Appendix \ref{appH}). Finally
$(h_\eps(\eps^{-1}x),m_\eps(\eps^{-1}x))\to (h(x),m(x))$ in sup-norm
as $\eps\to 0$,   $(h,m)$ the solution of the Stefan problem
\eqref{e2.14}.

             \end{thm}

\vskip.5cm

\noindent Theorem \ref{thme2.2} is proved in Section \ref{sec:e4} and in Appendix \ref{appDD},
\ref{appE},  \ref{appF}  and  \ref{appH} where we derive explicit bounds
on the speed of convergence.  By Theorem  \ref{thme2.1} we can
construct a quasi solution $(h_0,m_0)$ of \eqref{e2.22} with an error which around the interface $\eps^{-1}x_0$
is exponentially
small in $\eps^{-1}$ (we shall exploit this
with the introduction of suitable weighted norms).
$(h_0,m_0)$  is then used as the starting point of an iterative scheme similar to
the one in the proof of Theorem  \ref{thme2.1} from which however it differs significantly
due to the absence of symmetries.
The problem is that
we cannot restrict anymore to the space of
antisymmetric functions and thus need to
check that the maximal eigenvalue of the operator $L$
obtained by linearizing the first equation in  \eqref{e2.22} is non zero.
We know however from \cite{DOP} that it is close to zero and actually
vanishes as $\eps\to 0$.  But in our specific case
we can be more precise and
prove that it is negative and bounded away from 0 proportionally to $\eps$.  Thus we can invert
$L$ but get a dangerous factor $\eps^{-1}$ in the component
along the direction of the maximal eigenvector which
spoils the iterative scheme as it is and it thus needs to be modified.
The idea roughly speaking is to slightly shift
from $\eps^{-1}x_0$ to make smaller the component along the maximal eigenvector
(hence the condition $\dis{\int_{-\eps^{-1}\ell}^{\eps^{-1}\ell} h u^*=0}$ mentioned
before  Theorem \ref{thme2.2}) and this is
enough to make the iteration work.  The shifts described above are responsible for the delocalization of the
zero of the magnetization profile which may not coincide with that of the magnetic field.

\vskip1cm

\centerline{\em The Dirichlet problem.}
 \nopagebreak

\vskip.1cm  \noindent By Theorem \ref{thme2.1} and  \ref{thme2.2} it then follows
that there are solutions of the stationary mesoscopic equation which converge as
$\eps\to 0$ to the solution of any   Dirichlet problem with $m_-<-m_\beta$ and $m_+>m_\beta$
or viceversa.  At the mesoscopic level, though, the
boundary values may differ from the prescribed ones but the difference is infinitesimal in $\eps$.
We omit the proof that the above extends to any choice of $m_{\pm}$ in the complement of
$(-m_\beta,m_\beta)$ provided $m_+\ne m_-$.  We thus have a complete theory of the derivation
of the Stefan problem from \eqref{e2.22} gaining a deeper insight on the sense in which
the values in $(-m_\beta,m_\beta)$ are forbidden.  At the mesoscopic level in fact such a restriction
is absent and in the approximating profiles $(h_\eps,m_\eps)$ which at each $\eps$ solve \eqref{e2.22},
the values in  $(-m_\beta,m_\beta)$ are indeed present in
$m_\eps$.  However the fraction of space
where they are attained becomes negligible as $\eps\to 0$, they concentrate at the interface which in
macroscopic units becomes a point and in mesoscopic units are described to leading order by the instanton
which converges exponentially fast to $\pm m_\beta$.

\vskip1cm

\centerline{\em Under-cooling and over-heating effects.}
 \nopagebreak
\vskip.1cm  \noindent  In the forbidden interval $(-m_\beta,m_\beta)$ we distinguish two regions: one called
``spinodal'' is $[-m^*,m^*]$, $m^*=\sqrt {1-1/\beta}$, the other,
$\{m_\beta>|m|>m^*\}$, is called  metastable and it splits into
two disjoint intervals, the plus and minus metastable phases according to the sign of $m$.
In the spinodal region $\phi_\beta$ is concave, see \eqref{e2.2},
while in $(m^*,1)$ [as well as in $(-1,-m^*)$]
$\phi_\beta$ is strictly convex.
If we could restrict to  $(m^*,1)$ [or to  $(-1,-m^*)$]
ignoring or deleting the complement, then
$\phi_\beta$  would be convex and it could play the role of a thermodynamically
well defined
free energy giving rise to a new ``metastable thermodynamics'', new because
in the interval $(m^*,m_\beta)$ it  differs
from the ``true'' thermodynamic free energy $a_\beta$.
When (if ever) is it correct to use the
metastable one?  The usual answer (as its name suggests)
is that the time scale should not be too long and
the initial state of the system entirely in the plus [or in the minus] metastable phase.
When the evolution is given by \eqref{e2.18} initial states entirely in the plus phase
$(m^*,1)$ [or  in the minus one, $(-1,-m^*)$] evolve remaining in the plus [minus] phase,
so that the other values of the magnetization never enter
into play and can be ignored.  In particular if $m_\eps(x,t)$ solves \eqref{e2.18} with initial
datum $m_\eps(x,0)= m_0(\eps
x)$, $m_0\in C^\infty(\mathbb R^d;(m^*,1))$, then
   \begin{equation}
    \label{e2.24}
\lim_{\eps\to 0}
m_\eps(\eps^{-1}x,\eps^{-2}t) = m(x,t)
    \end{equation}
solution [with initial datum $m_0$] of
   \begin{equation}
    \label{e2.25}
\frac{\partial m}{\partial t}= {\rm div}\,\Big(D^*\,{\rm grad}\,m\Big),\quad D^*=1-\beta(1-m^2)
    \end{equation}
with $D^*>0$ in $(m^*,1)$ [and in $(-1,-m^*)$ as well].  By \eqref{e2.2}, $D^*=\chi \phi''_\beta$ which confirms
the interpretation of $\phi_\beta$ as a free energy once we compare $D^*$ with the expression for
$D_\beta$ in \eqref{e2.13}.  This is proved  in \cite{LOP} where
the analysis extends  to the
spin system with Kac potentials, if the Kac scaling parameter
is suitably related to $\eps$ so that the time scale is $\eps^{-2}$.  On much longer times,
which scale exponentially in $\eps^{-1}$,
large deviations and
tunnelling effects enter into play with the metastable phase becoming unstable,
see  \cite{asselah}.

All the above deals with  initial  states entirely in the plus [or in the minus] phase, much less
is known when they coexist.  A first answer is provided in this paper, see
Theorems \ref{thme2.1} and  \ref{thme2.2}, where however
the coexisting plus and minus phases
are the thermodynamically stable ones. In such cases
the whole interval $(-m_\beta,m_\beta)$
shrinks in the thermodynamic limit  to a point, not distinguishing between
metastable and spinodal values (thus in agreement with the macroscopic, thermodynamics of the model)
Our next theorem proves that there are also stationary solutions of
\eqref{e2.22} where
the plus
and minus metastable phases coexist.

\vskip.5cm

             \begin{thm}
        \label{thme2.3}
Let $j> 0$ then for any  positive $\ell$ smaller than some $\ell_j$,
there is
an antisymmetric pair $(h_\eps(x),m_\eps(x))$ which solves
the stationary
problem \eqref{e2.22} in $\eps^{-1}(-\ell,\ell)$ and such that
$(h_\eps(\eps^{-1}x),m_\eps(\eps^{-1}x))$ converges in sup norm
as $\eps\to 0$
to  $(h(x),m(x))$ solution of
the ``metastable'' Stefan problem:
   \begin{equation}
    \label{e2.26}
h^*(x)= \int_0^x \frac{ -j}{  \chi(m) },\qquad m=\phi_\beta'^{-1}(h^*)\quad \text{\rm in $(-\ell,\ell)\setminus \{0\}$}
        \end{equation}
$h_\eps$ is strictly decreasing while, to leading orders in $\eps$, $m_\eps$  first decreases
then increases (around the origin) and then again decreases.  The interval where
it increases has length $I_\eps$ and $\eps I_\eps\to 0$ as $\eps\to 0$.

             \end{thm}

\vskip.5cm

\noindent  The proof of Theorem \ref{thme2.3} is completely similar to the proof of
Theorem \ref{thme2.1} and it is therefore omitted.  We did not check that the result extends to
the   case $x_0\ne 0$.
The coexistence of the plus and minus  metastable phases is
related to the presence of a current which ``stabilizes'' the profile. If $j=0$ the stationary solution
would be close to an instanton except for boundary layers and in the thermodynamic limit
would converge to the Wulff shape which in this case is simply $m_\beta\; {\rm sign}(x)$.
We conjecture that the profiles described in  Theorem \ref{thme2.3} are ``metastable'' in the sense
that an additional noise at fixed  current $j$ would make the state  tunnel
toward the  solution with same $j$ described in Theorem \ref{thme2.1}.

\vskip2cm

\section{\hskip.2cm Proof of Theorem \ref{thme2.1}}
   \label{sec:e3}

In this section  we shall prove
Theorem \ref{thme2.1} which will be a corollary of
three theorems stated below and proved later
in three successive appendices.
For notational simplicity we suppose
$j<0$ and, as discussed in Remark (a) after Theorem
\ref{thme2.1}, we restrict to
odd functions, so that by default in
this section all functions are antisymmetric.
The analysis is based on an iterative scheme which is
outlined in the next two paragraphs. We shall define
a sequence $(h_n,m_n)$ which for each $n$ satisfies
the equality $m_n = \tanh\{\beta J^{\rm neum}*m_n+\beta h_n\}$
and prove that  $(h_n,m_n)$ converges as $n\to \infty$ in sup norm
to a limit $(h,m)$ which is the
desired solution of \eqref{e2.22}.

\vskip.5cm

\centerline {\em The starting element.}

\vskip.1cm
\nopagebreak
\noindent
We define  $h_0$ using \eqref{e2.20}
with  $m$ set equal to $m_0$, $m_0$ the
odd function defined for $x>0$ as
    \begin{equation}
    \label{e3.1}
m_0(x)=\bar m (x)\mathbf{1}_{[0,\xi_\eps]}(x)+
u(\eps [x-\xi_\eps])\mathbf{1}_{(\xi_\eps,\eps^{-1}\ell]}(x)
    \end{equation}
where:
$\bar m$ is the
instanton (see the paragraph {\it Instanton:
notation and properties} in Appendix \ref{appA});
$\xi_\eps= x_\eps+2n_0$,
$x_\eps:\;\bar m(x_\eps)= m_\beta -\eps$,
$n_0$  a large integer independent of $\eps$,
its value will be specified in the course of the proof of
Lemma \ref{lemmaeA.1}; as shown in Appendix \ref{appA}
$x_\eps$ scales as $\log\eps^{-1}$.
Finally, $u(r)$, $r\in [0,\ell-\eps \xi_\eps]$, is
the  solution of the macroscopic
equation \eqref{e2.14} (which in\eqref{e2.14} is denoted by $m$).
Since $h_0$ is obtained from $m_0$ by \eqref{e2.20}  then
   \begin{equation}
    \label{e3.3}
m_0 = \tanh\{\beta J^{\rm neum}*m_0 +\beta h_0\}
    \end{equation}
a property which
will be satisfied by all the elements of the sequence $(h_n,m_n)$.  Moreover, denoting by
$\|\cdot\|$ the sup-norm,
    \begin{equation}
    \label{eze}
 \sup_\eps \|m_0\| \le c_{\eqref{eze}} <1
    \end{equation}
because $\|\bar m\| \le m_\beta$ and $\|u\|<1$
since $\ell<\ell_j$, see
\eqref{e2.23} and the paragraph {\em  Axiomatic non equilibrium macroscopic theory}
in Section \ref{sec:e2}.

\vskip.5cm

\centerline {\em The iterative scheme.}

\vskip.1cm
\nopagebreak
\noindent
As discussed in Remark (a) after Theorem
\ref{thme2.1}, the idea is to
define a transformation
$h\to T(h)$ [from  antisymmetric into  antisymmetric
functions] in two steps. We first find an antisymmetric function $m$ such that
$m= \tanh\{\beta J^{\rm neum}*m+\beta h\}$ and then define for $x\ge 0$
    \begin{equation}
    \label{e3.4}
 T(h)(x)=-\eps j\int_0^x \chi(m(y))^{-1},\;\;\;  m= \tanh\{\beta J^{\rm neum}*m+\beta h\}
    \end{equation}
The definition of $T(h)$ thus rests on the possibility of finding an
``auxiliary function'' $m$ which solves
the second equality in \eqref{e3.4} and it is such that $\chi(m)^{-1}$ is integrable.
By construction we already know  that
the auxiliary function  $m_0$ associated to $h_0$ exists and $\|m_0\|\le c_{\eqref{eze}}<1$
uniformly in $\eps$.
The crucial step will
then be to prove that
if $h$ is ``close'' to $h_0$  then (at least for $\eps$ small enough)
there is a unique $m$ ``close'' to $m_0$
so that the second equality in \eqref{e3.4} is  satisfied,  $\|m\|<1$ and
$T(h)$ is thus
well defined (we do not have  general uniqueness as we are in the phase transition regime:
we cannot exclude that there are other
solutions not close to $m_0$). We  shall then prove recursively that all
images $h_n=T^n(h_0)$ are well defined and close to $h_0$, while the auxiliary functions $m_n$ are close
to $m_0$; moreover $(h_n,m_n)\to (h,m)$ in sup-norm
as $n\to \infty$.   $h$ will then  be
a fixed point of $T$ with auxiliary function
$m$ and Theorem \ref{thme2.1} will be proved.

\vskip.5cm

\centerline {\em  Notation.}

\vskip.1cm
\nopagebreak
\noindent
Our basic accuracy parameter will be
$\eps^a$, $a\in (0,1)$.
$\eps^a$ defines quantitatively the a-priori closeness to $h_0$ (the elements $h_k$ in the iteration
will actually be much closer to $h_0$, $\|h_k-h_0\|\le c \eps\log\eps^{-1}$):
    \begin{equation}
    \label{e3.5}
 \|h-h_0\|\le  \eps^a,\quad
\|f\|:= \sup_{|x|\le \eps^{-1}\ell} |f(x)|
    \end{equation}
being understood that all functions we deal with in this section
are odd.
While the
basic accuracy parameter clearly depends on $\eps$ , $a\in (0,1)$ above as well as
all the constants that we shall write
in the sequel, denoted by $a$, $b$, $c$ and  $C$ with or
without suffixes, will be independent of $\eps$. The existence of the auxiliary function $m$
in \eqref{e3.4} is established next:

\vskip.5cm

             \begin{thm}
        \label{thme3.1}
There are constants $c_{\eqref{e3.6}}>1$, $\alpha_{\eqref{e3.6}}>0$,  $c'_{\eqref{e3.6}}:=\frac{2c_{\eqref{e3.6}}}{\alpha_{\eqref{e3.6}}}$ so that
for all $\eps$ small enough the following holds.  For any $h:\|h-h_0\|\le \eps^a$
there is a unique $m_h$ in the ball $\{m:\|m-m_0\|\le c'_{\eqref{e3.6}} \eps^a\}$
such that $m_h=\tanh\{\beta J^{\rm neum}*m_h+\beta h\}$ and
for any $h':\|h'-h_0\|\le \eps^a$
    \begin{equation}
    \label{e3.6}
|m_h(x)-m_{h'}(x)| \le c_{\eqref{e3.6}}\int_{0}^{\eps^{-1}\ell}e^{-
\alpha_{\eqref{e3.6}}|x-y|} |h(y)-h'(y)|,\quad x\ge 0
    \end{equation}

             \end{thm}

\vskip.5cm

\noindent  We postpone to Appendix \ref{appA} the proof of
Theorem \ref{thme3.1} and proceed with the proof
of Theorem \ref{thme2.1} observing that as a consequence
of Theorem \ref{thme3.1} if $\|h-h_0\|\le\eps^a$ then $T(h)$
is well defined (for all
$\eps$ small enough)  because $\chi(m)$ in the first of \eqref{e3.4}
is bounded away from 0.
To prove this it suffices
to show that $\|m\|<1$.
By \eqref{e3.6} with $m_h=m$
and $m_{h'}=m_0$,
    \begin{equation}
    \label{e3.66}
\|m -m_{0}\| \le c'_{\eqref{e3.6}} \|h -h_0\|
\le c'_{\eqref{e3.6}} \eps^a
    \end{equation}
Then by \eqref{eze}
 $\|m\| \le c_{\eqref{eze}}+c'_{\eqref{e3.6}}\eps^a<1$  for $\eps$ small enough.

\vskip.5cm

             \begin{thm}
        \label{thme3.2}
There are constants $c_{\eqref{e3.7}}$ and $c_{\eqref{e3.8}}>0$ so that
for all $\eps$ small enough the following holds. Let $m'$
and $m''$ be both in the ball $\{m:\|m-m_0\|\le c'_{\eqref{e3.6}} \eps^a\}$, then
denoting by $\dis{h'=\int_0^x \frac{-\eps j}{\chi(m')}}$, $\dis{h''=\int_0^x \frac{-\eps j}{\chi(m'')}}$,
    \begin{equation}
    \label{e3.7}
|h'(x)-h''(x)| \le c_{\eqref{e3.7}}\eps|j|\int_{0}^{x}  |m'(y)-m''(y)|,\quad x> 0
    \end{equation}
    \begin{equation}
    \label{e3.8}
\|h_1-h_0\| \le c_{\eqref{e3.8}}\eps\log\eps^{-1},\quad h_1=T(h_0)
    \end{equation}

             \end{thm}

\vskip1cm

\noindent We  postpone to Appendix \ref{appB} the proof of Theorem \ref{thme3.2} and observe that
since the transformation $T$ is well defined in the ball $\|h-h_0\|\le \eps^a$  we are
in business once we show that any
iterate of $T$  is in the ball $\|h-h_0\|\le \eps^a$.
We  postpone to Appendix \ref{appC}
the proof of:

\vskip.5cm

             \begin{thm}
        \label{thme3.3}
There is a  constant $c_{\eqref{e3.9}}>0$ so that the following holds. Suppose there is $n$ such that
for all $k< n$, $h_k=T^k(h_0)$ is well defined,  $\|h_k-h_0\|\le \eps^a$ and
$\|m_k-m_0\|\le \eps^a$, $m_k$ the auxiliary function in
the definition of $T(h_k)$. Then $h_{n}$ is well defined and
    \begin{equation}
    \label{e3.9}
\|h_{k+1}-h_{k}\| \le c_{\eqref{e3.9}} (\frac 12)^k\|h_1-h_0\|,\;\;k< n
    \end{equation}

             \end{thm}

\vskip.5cm

\noindent  It is now easy to prove Theorem \ref{thme2.1}. We restrict  to $\eps>0$ so small that
    \begin{equation}
    \label{e3.6.2}
2c_{\eqref{e3.9}} c_{\eqref{e3.6.1}} c_{\eqref{e3.8}} \eps\log \eps^{-1} <\eps^a
    \end{equation}
(with $c_{\eqref{e3.6.1}}$ defined in \eqref{e3.6.1} below)
and prove by induction that
$(h_k,m_k)$ exists for all $k$ and moreover  $\|h_k-h_0\| \le \eps^a$ and
$\|m_k-m_0\|\le\eps^a$.  Since the statement is obviously true for $k=0$ we only
need to prove that if it is verified for $k<n$ then it holds for $n$ as well. By
\eqref{e3.9} and \eqref{e3.8} for all $k<n$,
    \begin{eqnarray}
 &&
\|h_{k+1}-h_{k}\| \le c_{\eqref{e3.9}}(\frac 12)^k c_{\eqref{e3.8}}\eps\log \eps^{-1}
     \label{e3.6.3i}
    \end{eqnarray}
which, by \eqref{e3.6.2} shows that $\|h_n-h_0\|<\eps^a$ (for $\eps$ small enough).  Then by Theorem \ref{thme3.1}
$m_n$ is well defined and by \eqref{e3.6} for all $k<n$
    \begin{equation}
    \label{e3.6.1}
\|m_{k+1} -m_{k}\| \le c_{\eqref{e3.6.1}}\|h_{k+1} -h_{k}\|,\quad
c_{\eqref{e3.6.1}}=\max\{1, c'_{\eqref{e3.6}} \}
    \end{equation}
Then, using \eqref{e3.6.3i},
    \begin{eqnarray}
&&\|m_{k+1}-m_{k}\| \le c_{\eqref{e3.9}}c_{\eqref{e3.6.1}}(\frac 12)^k  c_{\eqref{e3.8}}\eps\log \eps^{-1}
     \label{e3.6.3}
    \end{eqnarray}
which by \eqref{e3.6.2} proves that $\|m_n-m_0\|\le\eps^a$. Thus the induction is proved
and we   know that  for all $k$, $(h_k,m_k)$ exists,  $\|h_k-h_0\| \le \eps^a$ and
$\|m_k-m_0\|\le\eps^a$.

As a consequence of \eqref{e3.6.3} and \eqref{e3.6.3i}, $(h_n,m_n)\to (h,m)$ in sup-norm with $h=T(h)$,
$m=\tanh\{\beta J^{\rm neum}*m+\beta h\}$, and
    \begin{equation}
    \label{e3.9.1}
\|h-h_{0}\| \le c\eps\log\eps^{-1},\quad \|m-m_0\|\le c\eps\log\eps^{-1}
    \end{equation}
Making explicit the dependence on $\eps$ we write the limit as $(h_\eps,m_\eps)$ in agreement with
the notation in Theorem \ref{thme2.1}.  Recalling the definition of $(h_0,m_0)$, see \eqref{e3.1}, we
then obtain the proof of
Theorem \ref{thme2.1} except for the statement about the monotonicity of
$m_\eps$ which is proved at the end
of Appendix \ref{appDD}.

\vskip2cm

\section{\hskip.2cm Outline of the proof of Theorem \ref{thme2.2}}
   \label{sec:e4}

\centerline {\em The macroscopic solution.}

\vskip.1cm
\noindent
For the sake of definiteness
we suppose $j<0$ and $x_0>0$ and for notational simplicity
that the interval $(-\ell,\ell)$ is just the interval $(-1,1)$.
By assumption
 $(-1,1)$ is then strictly contained
in the  interval of length $2\ell_j$ and center $x_0$, i.e.\ the maximal interval
where the macroscopic
problem with parameters $(j,x_0)$ has solution
(see the paragraph {\em Axiomatic non equilibrium macroscopic theory} in Section \ref{sec:e2}).
We then
write $\ell^*=1+2x_0$ so that $x_0$ is the middle point of the interval $[-1,\ell^*]$ and,
for what said above,
$\ell^*+1<2\ell_j$ so that
the macroscopic problem has a  solution $(h_{\rm mac}(x),m_{\rm mac}(x))$, $x\in (-1,\ell^*)$ with
the following properties: it is a smooth pair of functions antisymmetric
around $x_0$ such that $\|m_{\rm mac}\|<1$ and
$\|h_{\rm mac}\|<\infty$ (so that $\inf \chi(m_{\rm mac})  >0$).

\vskip.5cm

\centerline {\em The pairs
$(h^*,m^*)$ and $(h_\eps,m_\eps)$.}
\vskip.1cm
\noindent
By  Theorem \ref{thme2.1}
for any $\eps>0$ small enough there is a pair
$(h^*(x),m^*(x))$, $x\in \eps^{-1}[-1,\ell^*]$  (dependence on $\eps$
is not made explicit) which solves  \eqref{e2.22} and is
antisymmetric around $\eps^{-1}x_0$.
Then there is $c_{\eqref{e4.0}}>0$ so that
   \begin{equation}
    \label{e4.0}
\beta \ge p_{h^*,m^*} \ge c_{\eqref{e4.0}}\quad \text{ for all $\eps>0$ small enough}
    \end{equation}
$\beta \ge p_{h,m}$ is true in general, see \eqref{e2.1222}; instead
$p_{h^*,m^*} \ge c_{\eqref{e4.0}}$ because by \eqref{e4.1.1}
$p_{h^*,m^*}=\chi(m^*)=\beta(1-(m^*)^2)$ and $\|m^*\|<1$ uniformly in $\eps$.  This
follows from the inequality  $\|m_{\rm mac}\|<1$ because, by  Theorem \ref{thme2.1},
$\lim_{\eps\to 0} \|m_{\rm mac}(x)-m^*(\eps^{-1}x)\|=0$.
We next define  $(h_\eps,m_\eps)$:
   \begin{eqnarray}
   \nn &&
m_\eps(x)= m^*(x),\;\;h_\eps(x)=h^*(x) + R_\eps(x),\;\;\;\; x\in \eps^{-1}[-1,1]
\\&&  R_\eps(x)= \int_{\eps^{-1}}^{\eps^{-1}+1}
J(x,y)[m^*(y)-m^*(2\eps^{-1}-y)]\,dy
 \label{e4.1}
    \end{eqnarray}
We have added the ``correction'' $R_\eps$ to have:
   \begin{equation}
    \label{e4.1a}
m_\eps= \tanh\{ \beta  [J^{\rm neum} *  m_\eps] +\beta h_\eps\}
    \end{equation}

    \vskip.5cm

  \begin{lemma}
  \label{lemma4.1}
There are  $r_{\eqref{4z.3}}>0$, $c_{\eqref{4z.3}}>0$ and $c'_{\eqref{4z.3}}>0$ so that
  \begin{equation}
    \label{4z.3}
\|\frac{dm^*}{dx}\| \le c'_{\eqref{4z.3}},
\quad \sup_{|x-\eps^{-1}x_0|\ge r_{\eqref{4z.3}}\log\eps^{-1}}|\frac{dm^*(x)}{dx}| \le
c_{\eqref{4z.3}} \eps
    \end{equation}
As a consequence $|R_\eps(x)| \le c
\eps \text{\bf 1}_{\eps^{-1}-1\le x \le \eps^{-1}}$.
    \end{lemma}

    \vskip.5cm

\noindent {\bf Proof.}
By differentiating the equality $m^* = \tanh\{\beta J^{{\rm neum},*}*m^*+\beta h^*\}$
(valid in the whole   interval $\eps^{-1}(-1,\ell^*)$, $J^{{\rm neum},*}$
the kernel with Neumann conditions at its endpoints)
we get
  \begin{equation*}
  \|\frac{dm^*}{dx}\|\le \beta (\|\frac{dJ^{{\rm neum},*}}{dx}\|\|m^*\| +\|\frac{dh^*}{dx}\|)\le c
     \end{equation*}
because
$\|dh^*/dx\|\le c \eps$ (as $h^*$ solves \eqref{e2.22}),
hence the first inequality in \eqref{4z.3}. The second one
is not as easy and it will be proved at the end of Appendix \ref{appDD}.  Using such inequality in \eqref{e4.1}
we readily see that $|R_\eps(x)| \le c
\eps \text{\bf 1}_{\eps^{-1}-1\le x \le \eps^{-1}}$, $c=c_{\eqref{4z.3}}$.  \qed

 \vskip.3cm

\noindent  Thus
$R_\eps$ is ``a small boundary field''  and except for the small error $R_\eps$,
$\chi(m_\eps) \frac{dh_\eps}{dx}=-\eps j$ so that
the pair $(h_\eps,m_\eps)$ is
``almost a solution'' of the stationary problem (which could be interpreted as
a true solution of a problem with suitably redefined
boundary conditions).

\vskip.5cm

\centerline {\em An interpolation scheme.}
\vskip.1cm
\noindent
A natural way to obtain a true solution from a quasi solution
is via
the implicit function theorem after writing \eqref{e2.22} as
a single equation $f(h,m)=0$ on the space of pairs of $L^\infty$ functions.
Unfortunately we do not have a good control of
the derivative of $f(h,m)$ which may in principle vanish.
The problem simplifies if we  try to solve only
the first one in \eqref{e2.22} and then use the second one to redefine
$h$, which opens the way to
an iterative scheme as the one used
in Section \ref{sec:e3}.
The crucial step is the following: find
$\tilde m$ such that
$\tilde m=\tanh\{\beta J^{\rm neum}*\tilde m+\beta \tilde h\}$
knowing $\tilde h$   and that  $\tilde h$  is ``close'' to
another field $\hat h$ for which there is $\hat m$ such that
$\hat m=\tanh\{\beta J^{\rm neum}*\hat m+\beta \hat h\}$.
To solve this problem we
interpolate writing $h(t)=t\tilde h+(1-t)\hat h$,
$t\in [0,1]$, and pretending that
for all $t$ there is $m(t)$ such that $m(t)=\tanh\{\beta J^{\rm neum}*  m(t)+\beta   h(t)\}$,
we differentiate and get an equation for $dm/dt$.  Its solution will then allow
to obtain $\tilde m$ as
$\dis{\tilde m= \hat m +\int_0^1 \frac{dm}{ds}\,ds}$.

\noindent
The main point in this procedure
is therefore the analysis of the equation for $dm/dt$.  This is \eqref{E.3} in Appendix \ref{appE},
here we just say that it has the form $\psi =(A_{h,m}-1)^{-1}\phi$ ($\psi$ the unknown),
where $A_{h,m}=p_{h,m}J^{\rm neum}*$, $p_{h,m}$
as in \eqref{e2.1222} (and $p_{h,m}=\chi(m)=\beta(1-m^2)$
because $m=\tanh\{\beta J^{\rm neum}*m+\beta h\}$),
$J^{\rm neum}*$ is the convolution operator with kernel $J^{\rm neum}$.  The non linearity of the problem
reflects in the fact that $(h,m)$ above is actually $(h(t),m(t))$ which is itself unknown but
the whole problem boils down to an accurate analysis of
the operator $A_{h,m}$ in a suitably large set of pairs $(h,m)$ (the set $\mathcal A$
in Appendix \ref{appDD}).
The same problem has appeared in the proof of Theorem \ref{thme3.1}, where however
we had the great simplification
to restrict
to
the space of antisymmetric functions. In such a restricted space
$\|A_{h,m}^{n_0}\| <1$
for a suitable integer $n_0$ uniformly in $\eps$ (see  Appendix \ref{appA}).
$(1 - A_{h,m})^{-1}$ is then equal
to the convergent sum $\sum A_{h,m}^n$ and the bound \eqref{eA.10} holds.
In the case
considered in  Theorem \ref{thme2.2} we do not have symmetries
and the invertibility of $L_{h,m}:= A_{h,m}-1$ becomes a serious issue.

In  Appendix \ref{appDD} we shall establish fine spectral properties of
$A_{h,m}$
for
all $(h,m)$ in a set  $\mathcal A$.
We shall prove
a Perron-Frobenius
theorem for
$A_{h,m}$
regarded as an integral operator
on $L^\infty(\eps^{-1}[-1,1])$ showing
that it has a maximal eigenvalue
$\la>0$, that its eigenvector $u$  (called the ''maximal eigenvector'')
has a definite sign (taken positive)
and, see Proposition \ref{propDD.1}, that there are positive constants
$c$, $ c'$ and $a$ so that
for all $\eps$ small enough
    \begin{equation}
    \label{e4.4}
0<\la< 1-c  \eps,\quad 0<u \le c' e^{- a|x-x_0|}
    \end{equation}
Actually to leading order in $\eps$, $\la= 1- C_{\eqref{DD.5}}\eps$, see \eqref{DD.5}.
$\la$ is separated from the rest of the spectrum (spectral gap) as stated
in Proposition \ref{propDD.2}.

Our strategy therefore will be to reduce to pairs $(h,m)\in \mathcal A$, a task accomplished by
showing that we can actually reduce to functions $h$ in the very small neighborhood
 $\mathcal G$  of $h_\eps$  defined next.

\vskip.5cm

\centerline{\em The  set  $\mathcal G$}
  \nopagebreak
   \vskip.1cm\noindent
Let $b_{\eqref{e4.8az}}$ and $a_{\eqref{e4.8a}}$
be positive parameters (specified in Appendix \ref{appF}), and
for any $f\in L^\infty(\eps^{-1}[-1,1])$
        \begin{equation}
    \label{e4.8a}
N(f):= \sup_{|x| \le  \eps^{-1}} E_\eps(x)|f(x)|;\;\;\; E_\eps(x):= \begin{cases} e^{a_{\eqref{e4.8a}}(\eps^{-1}-x)} &
x \ge \eps^{-1}x_0\\  e^{a^-_{\eqref{e4.8a}}(x+\eps^{-1})} & x < \eps^{-1}x_0 \end{cases}
    \end{equation}
with $a^-_{\eqref{e4.8a}}$ such that $a^-_{\eqref{e4.8a}}(x_0+1)=a_{\eqref{e4.8a}}(1-x_0)$.
Recalling that  $h_\eps$ is defined in \eqref{e4.1} and denoting by $u^*\in L^\infty(\eps^{-1}[-1,\ell^*],\mathbb R^+)$
the `` maximal eigenvector'' of $A_{h^*,m^*}$ we define $\mathcal G$ as
   \begin{eqnarray}
    \label{e4.8az}
\mathcal G:=\Big\{h&:& N(h-h_\eps) \le  b_{\eqref{e4.8az}},\;\; \int_{-\eps^{-1}}^{\eps^{-1}} h u^* =0\nn \\
&&  \|\frac{d(h-h_\eps)}{dx}\| \le \eps\;\;
\sup_{|x-\eps^{-1}x_0|\le (\log\eps^{-1})^2}|\frac{d(h-h_\eps)}{dx}|\le \eps^2 \Big\}
     \end{eqnarray}

\vskip.5cm

\centerline{\em The iterative scheme.}
\vskip.1cm
\noindent
We shall prove in  Proposition \ref{propF.1} that if $h\in \mathcal G$ then there is  $m$ such that
$m=\tanh\{\beta J^{\rm neum}*m+\beta h\}$ and moreover
$(h,m)\in \mathcal A$, $\mathcal A$ the nice set with good spectral
properties mentioned earlier. Thus $A_{h,m}$ has
a maximal eigenvalue  $\la$ with maximal eigenvector $u$,
$A_{h,m}u=\la u$.  In  Corollary \ref{coroF.1} we shall prove that $u$ is ``very close''
to the restriction of $u^*$ to $\eps^{-1}[-1,1]$, $u^*$ the maximal
eigenvector of $A_{h^*,m^*}$ relative to the problem in $\eps^{-1}[-1,\ell^*]$.
All this collects the properties needed to define the iterative scheme and
to prove its convergence.
  We define recursively $h_{n+1}:=T(h_n)$, $n\ge -1$,  $h_{-1}:=h_\eps$,  as
    \begin{equation}
    \label{e4.6}
h_{n+1} = \hat h_{n+1} - \frac{ \dis{\int_{-\eps^{-1}}^{\eps^{-1}} \hat h_{n+1} u^*}}
{\dis{\int_{-\eps^{-1}}^{\eps^{-1}}  u^*}},  \quad
 \hat h_{n+1}(x):=-\eps j\int_{\eps^{-1}x_0}^x \chi(m_n(y))^{-1}
    \end{equation}
(recalling that $\chi(m_n)=p_{h_n,m_n}$ by \eqref{e4.1.1}). The
definition is well posed once we prove that  $h_n\in \mathcal G$ for
$n\ge 0$, so that there is a  unique  $m_n$ such that $(h_n,m_n)\in
\mathcal A$. We shall indeed prove in Proposition \ref{propH.4} that
$N(h_{n+1}-h_n)\le c\eps N(h_n-h_{n-1})$.  Here we use in an
essential way the subtraction in \eqref{e4.6} which subtracts [most
of] the component along the maximal eigenvector $u$ of
$A_{h_{n-1},m_{n-1}}$ of the ``forcing term''
$p_{h_{n-1},m_{n-1}}(h_n-h_{n-1})$. In this way we shall prove
iteratively that $h_{n}\in \mathcal G$ so that there is $m_{n}$ with
$(h_n,m_n)\in \mathcal A$; moreover we shall see in Appendix
\ref{appH} that $h_n\to h$ and $m_n\to m$ as $n\to \infty$ with
$m=\tanh\{\beta J^{\rm neum}*m+\beta h\}$, $\dis{h = \hat h  -
\frac{ \int \hat h  u^*}{\int u^*}}$, $\hat h (x):=-\eps
j\int_{\eps^{-1}x_0}^x \chi(m(y))^{-1} $. As a consequence the pair
$(h,m)$ satisfies \eqref{e2.21} with $h(x_\eps)=0$ where $ x_\eps$
is such that:
    \begin{equation}
    \label{e4.8}
 \int_{\eps^{-1}x_0}^{x_\eps} \chi(m(y))^{-1} = \frac{ \int  u^*(x)
 \int_{\eps^{-1}x_0}^{x} \chi(m(y))^{-1}}{\int u^*}
    \end{equation}
The proof of Theorem \ref{thme2.2} will then be
completed by showing at the end of Appendix \ref{appH} that $x_\eps$ exists and
that $\eps x_\eps \to x_0$ as $\eps\to 0$, see \eqref{eG.26}.

\vskip2cm

\appendix

\section{\hskip.2cm Proof of Theorem \ref{thme3.1}}

\label{appA}

Before proving Theorem \ref{thme3.1} we introduce some notation
and definitions which will be used throughout
the whole sequel.

    \vskip.5cm
\centerline{ \em An auxiliary dynamics}
  \nopagebreak
    \vskip.1cm
  \noindent
  To construct and compare solutions of $m=\tanh\{\beta J^{\rm neum}*m+\beta h\}$ for given $h$,
we introduce an artificial dynamics.
Suppose   $(h(t),m(t))$, $t\in[ 0,1]$,  are smooth
functions of $t$ and that  for all $t$
   \begin{equation}
    \label{E.0}
    m(t)=\tanh\{\beta J^{\rm neum}*m(t)+\beta h(t)\}
        \end{equation}
        By differentiating \eqref{E.0} with respect to $t$ we get the identity
   \begin{equation}
    \label{E.1}
\frac{dm}{dt} = A_t \frac{dm}{dt}  + p_t
\frac{dh}{dt},\quad L_t \frac{dm}{dt}= - p_t
\frac{dh}{dt},\qquad L_t=A_t-1
    \end{equation}
where $p_t =p_{h(t),m(t)}$, $p_{h,m}$ as in \eqref{e2.1222}, and
$A_t=p_t J^{\rm neum}*$, $J^{\rm neum}*$
the operator on $L^\infty(\eps^{-1}[-\ell,\ell])$
with kernel $J^{\rm neum}$.

\noindent By a change of perspective we now
regard \eqref{E.1} as an equation for the unknown
$\frac{dm}{dt}$ with $p_t$ and $\frac{dh}{dt}$  considered as ``known
terms''.  We shall prove in this appendix that under suitable assumptions
on $h$
a solution  exists and it is
unique.  We then ``construct''  $\dis{m(t):= m(0)+\int_0^t \frac{dm}{ds}}$ and check that it
verifies
\eqref{E.0}.  The important point is that the whole procedure works in the same way even if we  ask
that \eqref{E.0} holds only at time $t=0$, being a by-product of the analysis
that it remains valid for all $t\in [0,1]$.
In the actual applications  $m(0)=m_0$ is a given, known
function which solves $m_0=\tanh\{\beta J^{\rm neum}*m_0+\beta
h_0\}$,  $h(t)= h_1 t +(1-t)h_0$ with $h_0$ and $h_1$ also known and
$m(t)$ the unknown, in particular we are interested in its value
$m_1$ at time $t=1$ when $h(1)=h_1$. \eqref{E.1} then becomes a non linear
evolution equation and it  will be
crucial to prove first that $L_t$ is  invertible, so that
the equation can be written in normal form
   \begin{equation}
    \label{E.1.1}
 \frac{dm}{dt}=  L_t^{-1}\Big(- p_t
\frac{dh}{dt}\Big)
    \end{equation}
and then  that $ L_t^{-1}\big(- p_t
\frac{dh}{dt}\big)$ is a Lipschitz function of $m$ .

\vskip.5cm

\centerline{ \em The operator $A_{h,m}$}
  \nopagebreak
  \vskip.1cm

\noindent The whole analysis relies on properties of the spectrum
of the operator $A_{h,m}= p_{h,m}J^{\rm neum}*$
(called $A_t$ when
$(h,m)=(h(t),m(t))$ as above).
We shall study $A_{h,m}$ in a
$L^{\infty}(\eps^{-1}[-\ell,\ell])$ setting and since we want
to prove that $A_{h,m}-1$ is invertible it
is crucial to prove that $1$ is not in the spectrum of $A_{h,m}$.
Regarded
as an operator on $L^2\big(\eps^{-1}[-\ell,\ell], p^{-1}_{h,m}(x) dx\big)$,
$A_{h,m}$ is self-adjoint, it has a maximal eigenvalue
$\la_{h,m}$ which is positive and the corresponding eigenvector $u_{h,m}$, called the maximal
eigenvector, can and will be  chosen
as strictly
positive, see \cite{DOP}.  Further assumptions on $h$ and $m$ will allow to prove that
$\la_{h,m} \le 1-c\eps$, $c>0$, and that the rest of the spectrum is strictly below 1
uniformly in $\eps$.  The bound on $\la_{h,m}$ will not be used in this appendix, see the proof of
\eqref{eA.7} below.

\vskip.5cm

\centerline{\em Instanton: notation and properties}
  \nopagebreak
    \vskip.1cm
\noindent The instanton $\bar m$ is a solution of the local mean field equation
$\bar m(x)= \tanh\{\beta J*\bar m(x)\}$, $x\in \mathbb R$, with the following properties (see
 Section 8.1 and 8.2 of \cite{presutti}).
$\bar m(x)$ is a strictly increasing, antisymmetric function which converges
    to $\pm m_\beta$ as $x\to \pm \infty$, more
precisely there are $c_{\eqref{DD.1}}$ and $a_{\eqref{DD.1}}$ both positive
so that for all $x\ge 0$
   \begin{equation}
    \label{DD.1}
0<m_\beta- \bar m(x)\le c_{\eqref{DD.1}}e^{-a_{\eqref{DD.1}} x},\quad
\frac{d\bar m(x)}{dx}\le c_{\eqref{DD.1}}e^{-a_{\eqref{DD.1}} x}
    \end{equation}
We write
  \begin{eqnarray}
&& \bar p = \beta(1-\bar m^2),\;\; \bar A = \bar p J*,\;\;
 \bar m'  = \frac{d\bar m }{dx},\;\;{\tilde m}'=
\frac{\bar m'}{\langle(\bar m') ^2\rangle_\infty)^{1/2}}
    \label{E.1.2}
       \end{eqnarray}
where $\dis{\langle f\rangle_\infty=\int_{\mathbb R} f \bar p ^{-1}}$.
In \cite{DOP} and Section 8.3 in \cite{presutti} it is proved that there are $a_{\eqref{DD.17}}>0$
and $c_{\eqref{DD.17}}$ so that for any bounded function $f$
  \begin{equation}
    \label{DD.17}
\Big|\int \bar A ^n (x,y)  \tilde f(y)
 dy\Big| \le  \|\tilde f\|\;  c_{\eqref{DD.17}}
 e^{-a_{\eqref{DD.17}} n },\;\; \tilde f = f-  \langle f \tilde m' \rangle_{\infty}
\tilde m'
    \end{equation}

\vskip1cm

\noindent We can now turn to the proof of Theorem \ref{thme3.1} and
restrict hereafter in this appendix
to the space
of   antisymmetric functions.   After
observing that by \eqref{DD.1}
    \begin{equation}
    \label{eA.6}
 x_\eps \le c \log\eps^{-1}
    \end{equation}
we  complete  the definition \eqref{e3.1} of $m_0$ by
fixing the integer $n_0$, chosen so that
    \begin{equation}
    \label{eA.7}
\|\bar A^{n_0}\psi\| \le e^{-a_{\eqref{eA.7}}} \|\psi\|,\quad a_{\eqref{eA.7}}>0
    \end{equation}
where $\psi$ above is any bounded antisymmetric function,
recall that $\|f\|$
denotes the sup norm of $f$.
Existence of $n_0$ follows from \eqref{DD.17} because
$\bar m'$ and $\bar p_{x_0}$ are symmetric
and $\psi$ antisymmetric so that
$\langle \psi \tilde m'\rangle_{\infty}=0$.

\vskip1cm

             \begin{lemma}

        \label{lemmaeA.1}
There is $a_{\eqref{eA.4}}>0$ so that
for any $c$, $a>0$  and all $\eps$ small enough
    \begin{equation}
    \label{eA.4}
\|A^{n_0}_{h,m}\psi\|\le e^{-a_{\eqref{eA.4}}}\|\psi\|,\quad \text{if $\|m-m_0\| \le  c \eps^a$,
$\|h-h_0\|\le c\eps^a$}
    \end{equation}
for any bounded odd function $\psi$.

             \end{lemma}

\vskip.5cm

\noindent {\bf Proof.}  As we shall see \eqref{eA.4} is a straight consequence of \eqref{eA.7} and of
  \begin{equation}
    \label{eA.8bb}
  \|\frac{p_{h,m} }
 { p_{h_0,m_0}}-1\|\le c'\eps^a
        \end{equation}
which follows directly from  \eqref{eze}  and the assumptions on $h$ and $m$.
We distinguish
``small'' and ``large'' values of $x_0$ in $A^{n_0}_{h,m}\psi(x_0)$.

\smallskip

{\it (i). {$x_0\in [0,x_\eps+n_0]$}}.  We write
   \begin{equation}
 A^{n_0}_{h,m}\psi(x_0)=\int  \psi(x_{n_0})  \prod_{k=1}^{n_0}
 \{  p_{h_0,m_0}(x_{k-1})J^{\rm neum}(x_{k-1},x_k) \frac{p_{h,m} (x_{k-1})}
 { p_{h_0,m_0}(x_{k-1})}\}  dx_1\cdots dx_{n_0}
    \label{eA.8}
        \end{equation}
Since $J^{\rm neum}$ has range 1,
$|x_i| \le  x_\eps+2n_0$ for all
$i=1,\dots n_0$. Then by \eqref{eA.6} for $\eps$ small enough,
$J^{\rm neum}(x_i,x_{i+1})=J(x_i,x_{i+1})$.  Moreover $p_{h_0,m_0}(x_i)=\bar p(x_i)$
(because $m_0(x)=\bar m(x)$,
$h_0(x)=0$ for $|x| \le x_\eps+2n_0$ ).
Thus by \eqref{eA.8bb}
        \begin{equation*}
\Big|A^{n_0}_{h,m}\psi(x_0)-\bar A^{n_0}\psi(x_0)\Big|\le c' n_0 \eps^a \|\psi\|
    \end{equation*}
and using \eqref{eA.7}, for all $\eps$ small enough
        \begin{equation*}
\Big|A^{n_0}_{h,m}\psi(x_0)\Big|\le e^{-a_{\eqref{eA.7}}}  \|\psi\| +
c' n_0 \eps^a \|\psi\| \le e^{-a_{\eqref{eA.4}}}\|\psi\|
    \end{equation*}

\vskip.5cm

{\it (ii). { $x_0\in [x_\eps+n_0,\eps^{-1}\ell]$}}. We then write
  \begin{equation}
 A^{n_0}_{h,m}\psi(x_0)=\int  \psi(x_{n_0})  \prod_{k=1}^{n_0}
 \{  p_{h,m}(x_{k-1})J^{\rm neum}(x_{k-1},x_k)  dx_1\cdots dx_{n_0}
    \label{eA.8bbb}
        \end{equation}
and since $J^{\rm neum}$ has
range 1, $x_i \ge x_\eps$ in  \eqref{eA.8bbb} for all $i=1,\dots n_0$.
When $x_i\in [x_\eps,\xi_\eps]$, $p_{h_0,m_0}(x_i)= \bar p(x_i) \le \beta(1-\bar m(x_\eps)^2)$
and by the definition of $x_\eps$, $\bar m(x_\eps) = m_\beta-\eps$.  Hence if $b'$ is such that
 $\beta(1-m_\beta^2)<b'<1$ then for all
$\eps$ small enough, $p_{h_0,m_0}(x_i) \le b'<1$.  When $x_i>\xi_\eps$, $p_{h_0,m_0}= \beta(1-u^2)$ and
since $u\ge m_\beta$, $p_{h_0,m_0}(x_i) \le \beta(1-m_\beta^2)<b'<1$.  Thus by  \eqref{eA.8bb}
$p_{h,m}(x_i)\le b<1$
$|A^{n_0}_{h,m}\psi(x_0)  |\le  b^{n_0}\|\psi\|$.
\qed

\vskip1cm
\noindent
By \eqref{eA.4}, $L_{h,m}=A_{h,m}-1$ is invertible and
    \begin{equation}
    \label{eA.10}
L_{h,m}^{-1}=- \sum_{n=0}^\infty A_{h,m}^n,\;\; \|L_{h,m}^{-1}\| \le \frac{c_{\eqref{eA.10}}}{1-a_{\eqref{eA.4}}}
    \end{equation}
where    $c_{\eqref{eA.10}}$ bounds $\sum_{n=1}^{n_0} \|A^n_{h,m}\|$.
Moreover:

\vskip1cm

             \begin{lemma}

        \label{lemmaeA.2}
There exist  $\alpha_{\eqref{e3.6}}>0$, (which defines
the parameter introduced in \eqref{e3.6}),
$c_{\eqref{eA.11}}$ and  $c_{\eqref{eA.12}}$, both larger than
$\dis{\max\{1,\frac{c_{\eqref{eA.10}}}{1-a_{\eqref{eA.4}}}\}}$,
so that
for any $c$ and all $\eps$ small enough
    \begin{equation}
    \label{eA.11}
|L_{h,m}^{-1} \psi(x)| \le c_{\eqref{eA.11}} \int_0^{\eps^{-1}} e^{-
\alpha_{\eqref{e3.6}}|x-y|} |\psi(y)|, \quad \|m-m_0\| \le  c \eps^a,\;
\|h-h_0\|\le c\eps^a
    \end{equation}
 for   any $x\ge 0$. Moreover  if also
$m': \|m'-m_0\|\le c\eps^a$, then
    \begin{equation}
    \label{eA.12}
\|L_{h,m}^{-1} -L_{h,m'}^{-1}| \le c_{\eqref{eA.12}} \|m-m'\|
    \end{equation}

             \end{lemma}

\vskip.5cm

\noindent {\bf Proof.} To prove \eqref{eA.11} we write $A_{h,m}^n(x,y)$ as the kernel of $A_{h,m}^n$ and have by \eqref{eA.10}
    \begin{equation*}
L_{h,m}^{-1} \psi(x) = - \sum_{n=0}^\infty \int A_{h,m}^n(x,y) \psi(y)=  - \int  \sum_{n=n(x,y)}^\infty
A_{h,m}^n(x,y) \psi(y)
    \end{equation*}
where $n(x,y)\ge |y-x|$ because $A_{h,m}(x,y)=p_{h,m}(x)J^{\rm neum}(x,y)$ is supported by $|x-y|\le 1$.
 \eqref{eA.11} then follows from
\eqref{eA.4}. To prove
\eqref{eA.12} we write
        \begin{equation*}
 L_{h,m}^{-1} -L_{h,m'}^{-1}= L_{h,m}^{-1}\big(A_{h,m}-A_{h,m'}\big) L_{h,m'}^{-1}
    \end{equation*}
use \eqref{eA.10} and that    $\|A_{h,m}-A_{h,m'}\|\le c  \|m-m'\|$.  \qed

\vskip1cm

\noindent  We shall study
 \eqref{E.1.1} with
     \begin{equation}
    \label{eA.14}
h(t)= t h''+(1-t) h', \;\; \text{\rm  $h'$ and $h''$ in the ball
$\|h-h_0\|\le \eps^a$}
    \end{equation}
so that $\|h(t)-h_0\|\le \eps^a$ and
$\dis{\|\frac{dh(t)}{dt}\|\le 2 \eps^{a}}$.
The initial datum  $m'$ is chosen so that $m'=\tanh\{\beta J^{\rm neum}*m'
+\beta h'\}$ and $\|m'-m_0\|\le c' \eps^{a}$
where $c':=\beta c_{\eqref{eA.11}}$.
To prove existence of solutions of \eqref{E.1.1}  we need to control the ``velocity field''
    \begin{equation}
    \label{eA.13}
V(h,m,\dot h) = - L_{h,m}^{-1} \big(p_{h,m} \dot h\big)
    \end{equation}
where $m$, $h$, $\dot h$ are antisymmetric functions.
To this end
we specify the ``free parameter'' $c$ which appears in the previous two lemmas
so that $c> 3c'$, $c':=\beta c_{\eqref{eA.11}}$.

\vskip1cm

             \begin{lemma}

        \label{lemmaeA.3}
For all $\eps$ small enough
the Cauchy problem in the interval $t\in [0,1]$
    \begin{equation}
    \label{eA.15}
 \frac{dm(t)}{dt} = V\big(h(t),m(t),\frac{dh(t)}{dt}\big),\quad m(0)=m'
    \end{equation}
has a unique solution $m(t)$ such that $\|m(t)-m_0\|\le 3c'\eps^a$.  Moreover,
$m(t)=\tanh\{\beta J^{\rm neum}*m(t) +\beta h(t)\}$ for all  $t\in [0,1]$.

             \end{lemma}

\vskip.5cm

\noindent {\bf Proof.} When  $\|m-m_0\|\le  3c' \eps^{a}$,
$\|h-h_0\|\le \eps^a$ the velocity field $V(h,m,\dot h)$ is bounded
(by \eqref{eA.10}) and Lipschitz (by \eqref{eA.12}), recall that by
\eqref{eA.14} $\|\dot h\|\le 2  \eps^a$. We thus have local
existence and uniqueness till when $\|m-m_0\|\le  3c' \eps^{a}$.
Till this time
    \begin{equation*}
\| \frac{dm(t)}{dt} \|\le \beta \|L_t^{-1} \frac{dh(t)}{dt}\| \le \beta
\frac{c_{\eqref{eA.10}}}{1-a_{\eqref{eA.4}}} 2\eps^a \le 2\beta c_{\eqref{eA.11}} \eps^{a}
    \end{equation*}
(recalling from Lemma \ref{lemmaeA.2} that
$\dis{c_{\eqref{eA.11}}\ge
\max\{1,\frac{c_{\eqref{eA.10}}}{1-a_{\eqref{eA.4}}}\}}$). Hence
$\|m(t)-m(0)\| \le 2c'\eps^a$ which ensures existence till $t=1$.
Recalling \eqref{eA.13} we get from \eqref{eA.15} that
    \begin{equation*}
 \frac{d}{dt} \Big( m(t)- \tanh\{\beta J^{\rm neum}*m(t) +\beta h(t)\}\Big)=0
    \end{equation*}
$m(t)- \tanh\{\beta J^{\rm neum}*m(t) +\beta h(t)\}$ is thus constant and being 0 initially
it is 0 at all times.  \qed

\vskip1cm

\noindent By taking $h'=h_0$ in \eqref{eA.14} by
Lemma \ref{lemmaeA.3} we conclude that for any $h:\|h-h_0\|\le
\eps^a$ there is $m$ which satisfies $m=\tanh\{\beta J^{\rm neum}*m  +\beta h \}$
and $\|m-m_0\|\le 3c'\eps^a$, $c'=\beta c_{\eqref{eA.11}}$.

\noindent
Finally to prove \eqref{e3.6} we write $h(t)=th+(1-t)h'$ so that
    \begin{equation*}
m_h -m_{h'}= \int_0^1 \frac{dm(t)}{dt}= -  \int_0^1
L_{t}^{-1} p_{t} ( h - h')
    \end{equation*}
and  \eqref{e3.6} follows from \eqref{eA.11}.  The proof of Theorem \ref{thme3.1} is complete.

\vskip2cm

\section{\hskip.2cm Proof of Theorem \ref{thme3.2}}

\label{appB}

By \eqref{eze} there is $b<1$ so that for all $\eps$ small enough  $\|m\| \le b<1$
in the ball
$\{m:\|m-m_0\|\le c_{\eqref{e3.6}} \eps^a\}$;
\eqref{e3.7} then
readily follows. To prove  \eqref{e3.8} we observe that $h_0(x)=0$ for
$x\in[0,x_{\eps}+2 n_0]$ because in such interval $m_0=\bar m$ and
$\bar m=\tanh \{\beta J^{\rm neum}\star \bar m\}$.  Thus, if $h_1=T(h_0)$ by \eqref{eA.6}
    \begin{equation*}
|h_1(x)-h_0(x)| \le c \eps\log\eps^{-1},\quad |x| \le \xi_\eps:=x_\eps+2n_0
    \end{equation*}
Define for $x>\xi_\eps$
        \begin{equation}
    \label{BB.2}
 h(x) =\int_{\xi_\eps}^x
\frac{-\eps j}{\chi(u)}=\int_{\xi_\eps}^x
\frac{-\eps j}{\chi(m_0)}
    \end{equation}
then
 $m_0(x)=u(\eps [x-\xi_\eps]) = \tanh \{\beta
u(\eps [x-\xi_\eps] +\beta h(\eps [x-\xi_\eps])\}$, hence
    \begin{equation}
 m_0(x)= \tanh\{\beta J^{\rm
neum}\star m_0(x)+\beta (u(\eps [x-\xi_\eps])-J^{\rm neum}\star
m_0(x)+ h(\eps [x-\xi_\eps]))\}
    \label{BB.1}
    \end{equation}
Since
$m_0(x)=\tanh\{\beta J^{\rm
neum}\star m_0(x)+\beta h_0( x \}$, by \eqref{BB.1}
    \begin{equation*}
|h_0(x)- h(\eps [x-\xi_\eps])| \le c\eps,\quad \text{and by \eqref{BB.2}}\;\;
|h_0(x) -\int_{\xi_\eps}^x \frac{-\eps j}{\chi(m_0)} | \le c \eps,\quad
    \end{equation*}
Since $\dis{h_1(x) =\int_{0}^x
\frac{-\eps j}{\chi(m_0)}}$, $\dis{|h_1(x) -h_0(x)| \le |h_0(x) -\int_{\xi_\eps}^x \frac{-\eps
j}{\chi(m_0)} | + c \eps \xi_\eps}$
hence \eqref{e3.8}.

\vskip2cm

\section{\hskip.2cm Proof of Theorem \ref{thme3.3}}

\label{appC}

By assumption for $x\ge 0$, $0\le m_k(x)\le  m_{0}(x)+\eps^a$, $k<n$. By
By \eqref{eze} $\|m_0\|<1$ so that
for all $\eps$ small enough,
$p_{h_{k},m_{k}}$ is uniformly bounded away from 0. There is therefore $C<\infty$ (recall
the current $j$ is a constant)  such that
    \begin{equation}
    \label{eC.1}
| h_{k+1}(x)-h_k(x)|\le C \eps\int_0^x |m_{k}(y)-m_{k-1}(y)|
    \end{equation}
By \eqref{e3.6} for any $y\in [0,\eps^{-1}]$,
    \begin{equation}
    \label{eeC.2}
|m_k(y)-m_{k-1}(y)| \le c \int_{0}^{\eps^{-1}\ell}e^{-
\alpha |y-z|} |h_k(z)-h_{k-1}(z)|
    \end{equation}
where we have dropped the suffixes from $c$ and
$\alpha$. We define $\psi_{k+1}(x)=
| h_{k+1}(\eps^{-1}x)-h_k(\eps^{-1}x)|$, $x\in [0,\ell]$ and by combining \eqref{eC.1}
and  \eqref{eeC.2} we get
    \begin{equation}
    \label{eeC.3}
\psi_{k+1}(x)\le c' \int_0^x dy \;\int_0^\ell e^{- \eps^{-1}
\alpha |y-z|}\psi_k(z)\eps^{-1} dz
    \end{equation}
Define $v_k(x)= e^{-bx}\psi_k(x)$, $b>0$ a large constant whose value will be specified later.
We have:
    \begin{equation}
    \label{eeC.4}
v_{k+1}(x)\le c' \int_0^x e^{-b(x-y)} dy \;\int_0^\ell e^{- \eps^{-1}
\alpha |y-z|+b(z-y)}v_k(z) \eps^{-1}dz
    \end{equation}
For $\eps$ so small that $\eps^{-1}\alpha >b$ we have
    \begin{equation}
    \label{eeC.5}
\|v_{k+1}\|\le c' \int_0^x e^{-b(x-y)} dy \; \frac{2\eps^{-1}} {\eps^{-1}\alpha -b}\|v_k\|
\le \frac{c'}{b}\frac{2\eps^{-1}} {\eps^{-1}\alpha -b}\|v_k\|
    \end{equation}
We choose $b$ so that $\dis{\frac{4c'}{\alpha b} =\frac 12}$.  Then
for all $\eps$ so small that $\dis{   \frac{\eps^{-1}} {\eps^{-1}\alpha
-b} \le \frac{2}{\alpha}}$
    \begin{equation*}
\|v_{k+1}\| \le \frac{1}{2}
\|v_k\|\;\;\text{which yields }\;\; \|\psi_{k+1}\| \le e^{b\ell} (
\frac{1}{2})^k \|\psi_1\|.
 \end{equation*}

\vskip1cm

\section{\hskip.2cm Spectral properties of $A_{h,m}$}

  \label{appDD}

In this appendix we shall first define a set $\mathcal A$ by weakening  properties of
the pair $(h_\eps,m_\eps)$ and then
prove spectral properties of
$A_{h,m}$ when $(h,m)$ is   in a small neighborhood of $\mathcal A$.

\vskip.3cm

\centerline{\em Instanton: additional notation}
  \nopagebreak
  \vskip.1cm
  \noindent
Referring to Appendix \ref{appA} for definition
and properties of the instanton $\bar m$, we denote by $\bar m_{x_0}$,
$x_0\in (-1,1)$, the translate of  $\bar m$ by $\eps^{-1}x_0$:
  \begin{equation}
\bar m_{x_0}(x)=
\bar m(x-\eps^{-1}x_0), \; \bar m'_{x_0} = \frac{d\bar m_{x_0}}{dx},\;
\bar p_{x_0}(x) = \beta(1-\bar m_{x_0}(x)^2),\; \bar A_{x_0}  := \bar p_{x_0} J*
    \label{DD.1a}
   \end{equation}

\vskip.3cm

\centerline{{\em Properties of the pair  $(h_\eps,m_\eps)$}}
  \nopagebreak
\begin{itemize}
\item $m_\eps=\tanh\{\beta J^{\rm neum}*m_\eps+h_\eps\}$, see \eqref{e4.1a}.

\item There are $r>0$ and $b>0$ so that $p_\eps(x) \le e^{-b}$ for all $|x-\eps^{-1}x_0| \ge r$.

\item $\dis{\|\frac{dm_\eps}{dx}\| <  c'_{\eqref{4z.3}}}$ (proved in Lemma
\ref{lemma4.1}) and for any $c>0$ there is $c'>0$
so that for all $\eps$ small enough
      \begin{equation}
         \label{zD.2}
             \sup_{|x- \eps^{-1}x_0| \le c \log \eps^{-1}} |m_\eps(x)-\bar m_{x_0}(x)| < c_{\eqref{zD.2}} \eps\log\eps^{-1}
   \end{equation}
because by \eqref{e3.9.1},
$\|m^*-m_0\|\le c\eps\log\eps^{-1}$.

\item  Since
$\dis{\frac{dh_\eps}{dx} = \frac{-\eps j}{ p_\eps(x)}}$ and   $\inf p_\eps>0$ then
$\| h_\eps \| \le c_1$,
$\dis{\|\frac{dh_\eps}{dx}\|  <  c\eps}$ and, by \eqref{zD.2},
  \begin{equation*}
 \sup_{|x- \eps^{-1}x_0| \le c \log \eps^{-1}} |\frac{dh_\eps(x)}{dx} - \frac{-\eps j}{\bar p_{x_0}(x)}| < c'_1 \eps^2\log\eps^{-1}
    \end{equation*}

\end{itemize}

\vskip.5cm

\centerline{{\em The set  $\mathcal A$}}
  \nopagebreak
  \vskip.1cm

  \noindent
By default all coefficients $a$, $c$, $C$ with or without a suffix are meant to be
positive  and independent of $\eps$; we shall indicate below by  item $n$
the $n$-th property of $(h_\eps,m_\eps)$ as listed in the previous paragraph and introduce the quantities (with $b$ in \eqref{DD.2} below the parameter entering in item 2)
   \begin{equation}
    \label{DD.2}
C_{\eqref{DD.2}}>1 \;:\; e^{-a C_{\eqref{DD.2}}(1-x_0) \log\eps^{-1}} = \eps^2,\quad a := \min\big\{\frac b4,
a_{\eqref{DD.17}},a_{\eqref{DD.1}} \big\}
    \end{equation}
   \begin{equation}
    \label{DD.00}
    I=\{x:|x-\eps^{-1}x_0|\le 2C_{\eqref{DD.2}} \log\eps^{-1}\},\quad   I'=\{x:|x-\eps^{-1}x_0|\le C_{\eqref{DD.2}} \log\eps^{-1}\}
        \end{equation}
($I'$ will be used later in Proposition \ref{propDD.1}).
With such notation we define
$\mathcal A$  as the collection of all pairs  $(h,m)$ such that
$m=\tanh\{\beta J^{\rm neum}*m+\beta h\}$ and the following three inequalities hold:
   \begin{eqnarray}
    \label{DD.4}
&&p_{h,m}(x)=\beta\big(1-m(x)^2\big) \le e^{-2a_{\eqref{DD.4}}},
  \quad |x-\eps^{-1}x_0| \ge r_{\eqref{DD.4}}
    \\
    \label{DD.3}
&&\|\frac{dm}{dx}\| \le C_{\eqref{DD.3}},\;\;
\sup_{x\in I}|m(x)-\bar m_{x_0}(x)| \le c'_{\eqref{DD.3}} \eps\log\eps^{-1}
    \\
    \label{DD.3a}
&& \| h \| \le C_{\eqref{DD.3a}},\;\; \|\frac{dh}{dx}\|  \le  C_{\eqref{DD.3a}}  ,\quad
\sup_{x\in I} |\frac{dh(x)}{dx} - \frac{-\eps j}{\bar p_{x_0}(x)}| \le c_{\eqref{DD.3a}} \eps^2\log\eps^{-1}
    \end{eqnarray}
where $C_{\eqref{DD.3a}}>2$ and:

\begin{itemize}

\item  $r_{\eqref{DD.4}}>r$ and $2a_{\eqref{DD.4}}= b/2$, $b$ and $r$ are as in item 2.

\item
$C_{\eqref{DD.3}}>2c'_{\eqref{4z.3}}$
and $c'_{\eqref{DD.3}}>2c_{\eqref{zD.2}}$ (see item 3).

\item    $C_{\eqref{DD.3a}}>2\max\{c,c_1\}$ and $c_{\eqref{DD.3a}} >2c_1'$ (see item  4).

\end{itemize}

\vskip.1cm

\noindent With the above
choice of   parameters  $(h_\eps,m_\eps) \in \mathcal A$.

\vskip.5cm

\centerline{{\em Spectral properties in a neighborhood of  $\mathcal A$}}
  \nopagebreak
  \vskip.1cm
  \noindent
We continue the analysis of the spectrum of $A_{h,m}$ started
in   Appendix \ref{appA}
assuming that $(h,m)$ is in the $\delta$ ball of $\mathcal A$
defined as $\dis{\bigcup_{(h,m)\in \mathcal A}
B_{\delta}(h,m)}$,  $B_{\delta}(h,m):=\{ (h',m'):\|h-h'\|\le \delta, \|m-m'\|\le \delta\}$.

\vskip.5cm

  \begin{prop}
  \label{propDD.1}
There are positive constants $C_{\eqref{DD.5z}}$,
$c_{\eqref{DD.5}} $, $c'_{\eqref{DD.7}}$,
$c_{\eqref{DD.7a}}$ and $a _{\eqref{DD.7a}}$
so that for any $\eps$ small enough there is $\delta=\delta(\eps)>0$
and  for any $(h,m)$ in the $\delta$ ball of $\mathcal A$
      \begin{equation}
    \label{DD.5z}
p_{h,m} \ge C_{\eqref{DD.5z}}
         \end{equation}
      \begin{equation}
    \label{DD.5}
|\la_{h,m} -[1- C_{\eqref{DD.5}}\eps] | \le c_{\eqref{DD.5}}  (\eps \log\eps^{-1})^2,\quad
C_{\eqref{DD.5}} = |j| \frac{\langle \bar m' \rangle_{\infty}}{\langle (\bar m')^2 \rangle_{\infty}} >0
         \end{equation}
Moreover, let $u_{h,m}>0$ be normalized as $\langle u_{h,m}^2\rangle_{h,m}=1$ and $I'$ is
as in \eqref{DD.00}, then
     \begin{eqnarray}
    \label{DD.7}
&&\sup_{x\in I'}| u_{h,m} (x) - {\tilde m}'_{x_0}(x)| \le   c'_{\eqref{DD.7}} \eps (\log\eps^{-1})^2
         \\
    \label{DD.7a}
&& u_{h,m} (x)   \le   c_{\eqref{DD.7a}} e^{-a _{\eqref{DD.4}}|x-\eps^{-1}x_0|}
         \end{eqnarray}

  \end{prop}

\vskip.5cm

\noindent {\bf Proof.}
We shall first prove with slightly
better coefficients the inequalities \eqref{DD.5z}--\eqref{DD.7a}  when  $(h,m)$ is in $\mathcal A$ and
then use a continuity argument to extend the analysis to a $\delta$ ball
of $\mathcal A$.
We thus fix $(h,m)\in \mathcal A$ and
drop the suffix $(h,m)$ when no ambiguity may arise.

\vskip.5cm

$\bullet$\; Proof of \eqref{DD.5z}. We bound
$|m(x)| \le \tanh\{ \beta J^{\rm neum}*\|m\|+\beta \|h\|\}$ and
$\|h\| \le C_{\eqref{DD.3a}}$,
hence  $p_{h,m} \ge 2C_{\eqref{DD.5z}}$,
with $2C_{\eqref{DD.5z}}= \beta(1-s^2)$,
$s$ the positive solution of
$s=\tanh\{\beta s +\beta C_{\eqref{DD.3a}}\}$. \eqref{DD.5z} then follows in a $\delta$ ball
of $(h,m)$ if $\delta$ is small enough.

\vskip.5cm

We shall next  prove
some rough bounds on $\la$ and $u$ which will then
be improved as required in the proposition. We take here
$(h,m)$ in a $\delta$-ball of $\mathcal A$ with $\delta$ small enough.  We are
going to use repeatedly
variants of
the obvious equality:
\begin{equation}
    \label{DD.8.1}
\langle f A_{h,m} g \rangle_{h,m} = \langle f A_{h',m'} g \rangle_{h',m'} = \int f J^{\rm neum}*g
    \end{equation}
We have the lower bound
$\dis{\la \ge  \frac{\langle  \bar m'_{x_0}  A  \bar m'_{x_0} \rangle_{h,m}}{ \langle ( \bar m'_{x_0} )^2 \rangle_{h,m}}}$, $A\equiv  A_{h,m} $ and $\bar m'_{x_0}$ here restricted to $\La = \eps^{-1}[-1,1]$. Using \eqref{DD.8.1} we can rewrite the numerator as
    \begin{eqnarray*}
\langle  \bar m'_{x_0}  A  \bar m'_{x_0} \rangle_{h,m} &=& \int_{\La\times \La}  \bar m'_{x_0}(x)J^{\rm neum}
(x,y)  \bar m'_{x_0}(y)=  \int_{\mathbb R\times \mathbb R}  \bar m'_{x_0}(x)J
(x,y)  \bar m'_{x_0}(y) +\Delta \nn\\&=& \int_{\mathbb R}  \bar m'_{x_0}(x)^2/\bar p_{x_0} +\Delta
= \int_{\La}  \bar m'_{x_0}(x)^2/\bar p_{x_0} +\Delta'
\nn\\&=& \langle ( \bar m'_{x_0} )^2 \rangle_{h,m} +
  \int_{\La}  \bar m'_{x_0}(x)^2 \frac{p-\bar p_{x_0}}{ p \,\bar p_{x_0}} +\Delta',\quad
  p\equiv p_{h,m}
         \end{eqnarray*}
where by \eqref{DD.1} and \eqref{DD.2}, $|\Delta|$ and  $|\Delta'|$ are
both bounded by $ \le c e^{-a_{\eqref{DD.1}} \eps^{-1}(1-x_0)}\le c \eps^2$.  The denominator in the last integral is bounded from below because $p\equiv p_{h,m} \ge C_{\eqref{DD.5z}}$, (\eqref{DD.5z} has already been proved) and $\bar p_{x_0} \ge \beta(1-m_\beta^2)$ (as $\bar m(x)$ converges monotonically to
$m_\beta$ as $x\to \infty$).  By \eqref{DD.3} and for $\delta$ small enough
$|p(x)-\bar p_{x_0}(x)| \le 2c'_{\eqref{DD.3}}\eps\log\eps^{-1}$ when  $x\in I$,
while in the complement we  bound  $\bar m'_{x_0}$
as in \eqref{DD.1} (recalling \eqref{DD.2}) and use that
$|p-\bar p_{x_0}|\le \beta$.
In conclusion we get
\begin{equation}
    \label{DD.8}
  \la  \ge 1 - c_{\eqref{DD.8}} \eps \log\eps^{-1}
    \end{equation}
with $c_{\eqref{DD.8}}$ dependent  on $C_{\eqref{DD.3a}},c'_{\eqref{DD.3}},a_{\eqref{DD.4}}$.
\vskip.5cm

$\bullet$\; Proof of  \eqref{DD.7a}.
We use \eqref{DD.8} and
the identity $ u (x)= \la ^{-n}(A ^n u )(x)$ to get   upper bounds on $u$.
With $n=1$ we obtain
  \begin{equation}
    \label{DD.16b}
\|u\|  \le \la^{-1}\|J\| \beta \sqrt 2
(\int  u^2)^{1/2}  \le \la^{-1}\|J\| \beta \sqrt 2
(\int \frac{\|p\|}{p} u^2)^{1/2}
\le c \langle u ^2 \rangle_{h,m} ^{1/2}
    \end{equation}
(having used Cauchy-Schwartz and that $\|p\|\le \beta$). By tuning $n$ with the
distance from  $\eps^{-1}x_0$ we get, using \eqref{DD.4},
    \begin{equation}
    \label{DD.14}
u (x) \le [1 - c_{\eqref{DD.8}} \eps \log\eps^{-1}]^{-n} e^{-2a_{\eqref{DD.4}} n}
\| u\|,\quad  \text{ when $|x-\eps^{-1}x_0|\ge n+r_{\eqref{DD.4}}$}
    \end{equation}
which together with \eqref{DD.16b}
proves  \eqref{DD.7a}
for $(h,m)\in \mathcal A$.

\vskip.5cm

\noindent
We shall next prove an upper bound on $\la\equiv \la_{h,m}$, $(h,m)$
in a $\delta$-ball of $\mathcal A$, $\delta$ suitably small.  We start
from the operator $\bar A_{x_0}= \bar p_{x_0} J*$ acting on $L^\infty(\mathbb R)$
and since 1 is its maximal eigenvalue (with eigenvector $\bar m'_{x_0}$),
$\dis{1 \ge  \frac{\langle u  \bar A_{x_0} u\rangle_{\infty}}
{ \langle  u^2 \rangle_{\infty}}}$ where we choose $u=u_{h,m}$ on
$\La = \eps^{-1}[-1,1]$ and $u=0$ on $\La^c$. Denoting $\dis{\langle f\rangle_\infty=
\int_{\mathbb R} \frac{f}{\bar p_{x_0}}}$, we then have
    \begin{equation*}
\langle u  \bar A_{x_0} u\rangle_{\infty}= \int_{\La\times \La} u(x) J(x,y)u(y) =
 \int_{\La\times \La} u(x) J^{\rm neum}(x,y)u(y) +R
         \end{equation*}
with $\dis{R = -\int _{\La\times \La^c} u(x) J(x,y) u(y_\La)}$,
$y_\La$ the reflection of $y$ into $\La$ through
its endpoints.  By \eqref{DD.7a} $|R|\le c e^{-a_{\eqref{DD.4}}|\eps^{-1}(1-x_0)}\|u\|^2$, hence
writing hereafter $\langle \cdot \rangle = \langle \cdot \rangle_{h,m}$,
    \begin{eqnarray*}
 \langle  u^2 \rangle_{\infty}  &\ge&\langle u  \bar A_{x_0} u\rangle_{\infty}\ge  \langle u  A u\rangle - c e^{-a_{\eqref{DD.4}}|\eps^{-1}(1-x_0)}\|u\|^2
\\&=&  \la \langle u^2\rangle - c e^{-a_{\eqref{DD.4}}|\eps^{-1}(1-x_0)}\|u\|^2
         \end{eqnarray*}
Thus, by \eqref{DD.16b}, $\dis{\la \le \frac{\langle u^2\rangle_\infty}{\langle u^2\rangle}
+ c e^{-a_{\eqref{DD.4}}|\eps^{-1}(1-x_0)}}$.  We postpone the proof that
  \begin{eqnarray}
  \label{DD.e2.1}
 \big| \frac{\langle u^2 \rangle}{\langle u^2 \rangle_{\infty}} - 1\big|
 \le c \eps^2 + c\eps\log\eps^{-1}
     \end{eqnarray}
and conclude, recalling  \eqref{DD.8} and pending the validity of \eqref{DD.e2.1},
   \begin{equation}
    \label{DD.13}
1 - c_{\eqref{DD.8}} \eps \log\eps^{-1}\le  \la  \le 1+
c_{\eqref{DD.13}} \eps \log\eps^{-1}
    \end{equation}
with $c_{\eqref{DD.13}}$ dependent  on $C_{\eqref{DD.3a}}, c'_{\eqref{DD.3}}, a_{\eqref{DD.4}}$.

 \vskip.5cm

\centerline{Proof of \eqref{DD.e2.1}}
 \nopagebreak
 \vskip.1cm
 \noindent
Recalling
that $p\ge C_{\eqref{DD.5z}}$, $p \le \beta$, $\bar p_{x_0} \ge \beta(1-m_\beta^2)$  and
$\bar p_{x_0}\le \beta$, we have
  \begin{equation}
    \label{zD.18}
 c_{\eqref{zD.18}}^{-1} \le \frac{\langle u^2 \rangle}{\langle u^2 \rangle_{\infty}} \le c_{\eqref{zD.18}}
    \end{equation}
Then, by \eqref{DD.7a}
  \begin{eqnarray}
    \label{DD.16aaa}
 \frac{u (y)}{\langle u^2 \rangle_{\infty}^{1/2}}
 \le
  c
 e^{-a_{\eqref{DD.4}}|y-\eps^{-1}x_0| },
 \quad  y\in \La\setminus I'
    \end{eqnarray}
hence by \eqref{DD.2} and since $\dis{a_{\eqref{DD.4}}= \frac b4}$,
  \begin{equation}
    \label{DD.e1}
\int_{\La\setminus I'} u^2/\bar p_{x_0} \le c_{\eqref{DD.e1}} \langle u^2 \rangle_{\infty}
\eps^2,\qquad  \langle u^2 \rangle_{\infty} > \int_{I'} u^2/\bar p_{x_0} \ge (1-c_{\eqref{DD.e1}}\eps^2)
 \langle u^2 \rangle_{\infty}
    \end{equation}
We also have
  \begin{eqnarray*}
&& \big| \frac{\langle u^2 \rangle}{\langle u^2 \rangle_{\infty}} - \frac{\int_{I'} u^2/p}{\langle u^2 \rangle_{\infty}}\big|
 \le c \eps^2,\quad
 \big| \frac{\langle u^2 \rangle}{\langle u^2 \rangle_{\infty}} - \frac{\int_{I'} u^2/\bar p_{x_0}}{\langle u^2 \rangle_{\infty}}\big|
 \le c \eps^2 + c\eps\log\eps^{-1} \nn\\&&
 \big| \frac{\langle u^2 \rangle}{\langle u^2 \rangle_{\infty}} - 1\big|
 \le c \eps^2 + c\eps\log\eps^{-1}
     \end{eqnarray*}
In the first inequality above we have used \eqref{DD.16aaa}, in
the second  \eqref{DD.3} and in the third \eqref{DD.e1}. \eqref{DD.e2.1} is proved.

\vskip.5cm

We shall next prove \eqref{DD.7} that we split in an upper and a lower bound for $u=u_{h,m}$,
we take here $(h,m)$ in a $\delta$-ball of $\mathcal A$ with $\delta$ small enough

\vskip.5cm

\noindent $\bullet$\; Proof of \eqref{DD.7} (the upper bound).
Let $y\in I'$, then, writing below $y_0\equiv y$,
\begin{equation}
    \label{DD.15}
\la ^n u (y)=  \int  u (y_n)
 \prod_{k=1}^n \{\bar A_{x_0}(y_{k-1},y_k) \frac{p (y_{k-1})}
 {\bar p_{x_0}(y_{k-1})}\}  dy_1\cdots dy_n
    \end{equation}
We choose again $n=C_{\eqref{DD.2}} \log\eps^{-1}$ observing
that since $y_0\in I'$ all $y_k$ are in $I$. We   bound $\la ^{-n} \le
(1 - c_{\eqref{DD.8}} \eps \log\eps^{-1})^{-n} \le (1+ n c \eps \log\eps^{-1}) \le
(1+  c' \eps [\log\eps^{-1}]^2)$.
Since all $y_k$ are in $I$, by \eqref{DD.3} and for $\delta$ small enough,
  \begin{equation*}
 \prod_{k=1}^n   \frac{p (y_{k-1})}
 {\bar p_{x_0}(y_{k-1})} \le 1+  c \eps [\log\eps^{-1}]^2
    \end{equation*}
hence (with a new constant $c$)
  \begin{equation}
    \label{DD.16}
 u (y)\le
  [1 +
 c \eps(\log\eps^{-1})^2]\int u (y_n)
 \prod_{k=1}^n \{\bar A_{x_0}(y_{k-1},y_k)\} \}  dy_1\cdots dy_n
    \end{equation}
We  define $\tilde u$ so
that $\dis{u (y_n)= \langle \tilde m'_{x_0} u \rangle_{\infty}\; \tilde m'_{x_0}(y_n)
 + \tilde u}$.
By  \eqref{DD.17}-\eqref{DD.2}  and for all  $y\in I'$
  \begin{equation}
    \label{DD.16a}
 u (y)\le  \tilde m'_{x_0}(y)
   [1 +
 c \eps(\log\eps^{-1})^2] \langle \tilde m'_{x_0} u \rangle_{\infty} + c\eps^2 \|u\|
    \end{equation}
which by \eqref{DD.16b} can be rewritten as
  \begin{equation}
    \label{DD.16aa}
 \frac{u (y)}{\langle u^2 \rangle^{1/2}}
  \le
 \{ [1 + c  \eps( \log\eps^{-1})^2] \frac{\langle \tilde m'_{x_0} u \rangle_\infty}
{\langle u^2 \rangle^{1/2}}  \}
   \tilde m'_{x_0}(y)
 +c  \eps^2
    \end{equation}
By Cauchy-Schwartz,
  \begin{eqnarray}
    \label{DD.e3}
 \frac{u (y)}{\langle u^2 \rangle ^{1/2}}
  \le (\frac{\langle u^2 \rangle_{\infty}}{\langle u^2 \rangle })^{1/2} \Big\{
   [1 + c  \eps( \log\eps^{-1})^2]
    \tilde m'_{x_0}(y)\Big\}
 +c  \eps^2
       \end{eqnarray}
which, by \eqref{DD.e2.1},  proves
  \begin{equation}
    \label{DD.e1.1}
   \frac{u (x)}{\langle u^2 \rangle ^{1/2}}   \le  {\tilde m}'_{x_0}(x)+
    \frac{c'_{\eqref{DD.7}}}2 \eps (\log\eps^{-1})^2
   \end{equation}

\vskip.5cm

\noindent $\bullet$\; Proof of \eqref{DD.7} (the lower bound).
Proceeding in a similar way we get the lower bound:
  \begin{equation}
    \label{DD.16bb}
 \frac{u (y)}{\langle u^2 \rangle^{1/2}}
 \ge
 \{ [1 - c  \eps( \log\eps^{-1})^2] \frac{\langle \tilde m'_{x_0} u \rangle_{\infty}}
{\langle u^2 \rangle^{1/2}}  \}
    \tilde m'_{x_0}(y)
  \; - c \eps^2
    \end{equation}
To bound the curly bracket from below we multiply both sides of \eqref{DD.16a}
by $\bar p_{x_0}^{-1} u$ and integrate
over $I'$. By \eqref{DD.e1}:
  \begin{equation*}
 (1-c_{\eqref{DD.e1}}\eps^2)
 \langle u^2 \rangle_{\infty} \le   \langle \tilde m'_{x_0} u \rangle_{\infty}^2
 [1 + c  \eps( \log\eps^{-1})^2] + c \eps^2  \log\eps^{-1}  \|u\|^2
    \end{equation*}
By \eqref{DD.16b}, $\dis{ (1-c_{\eqref{DD.e1}}\eps^2) \le  \frac{\langle \tilde m'_{x_0} u \rangle_{\infty}^2}
{\langle u^2 \rangle_{\infty} }
 [1 + c  \eps( \log\eps^{-1})^2] + c \eps^2\log\eps^{-1}}$, hence
      \begin{equation}
    \label{DD.20}
1 - c  \eps( \log\eps^{-1})^2 \le  \frac{\langle \tilde m'_{x_0} u \rangle_{\infty}^2}
{\langle u^2 \rangle_{\infty} }
\le 1
    \end{equation}
which by \eqref{DD.16bb} yields  $\dis{ \frac{u (y)}{\langle u^2 \rangle^{1/2}} \ge
 (\frac {\langle u^2 \rangle_{\infty}}{\langle u^2 \rangle})^{1/2} [1 - c  \eps( \log\eps^{-1})^2]
  \tilde m'_{x_0}(y)  -
 c  \eps( \log\eps^{-1})^2
}$. Using \eqref{DD.e2.1} we then get
  \begin{equation}
    \label{DD.e1.2}
    u_{h,m} (x) \ge  {\tilde m}'_{x_0}(x) -  \frac{c'_{\eqref{DD.7}}}2 \eps (\log\eps^{-1})^2
   \end{equation}

\vskip.5cm

\noindent $\bullet$\; Proof of \eqref{DD.5}.  We first suppose $(h,m)\in \mathcal A$ and use for the first time the conditions on $dm/dx$ and $dh/dx$ contained in the definition of
$\mathcal A$.  Writing $f'$ for  the derivative of $f$
w.r.t.\ $x$,
we differentiate  $m(x)=\tanh\{\beta J^{\rm neum}*m(x)+\beta h(x)\}$ and
get
$m' =p J^{\rm neum}*m' +p  h'$, hence
$L m' = - p   h'$,  $L =A -1$.
We multiply both sides by $p^{-1}u$ and integrate over $x$. Recalling that $L$
is selfadjoint in the scalar product with weight $p^{-1}$, we then have
     \begin{equation}
    \label{eD.6}
(\la -1) \langle u  m' \rangle   = -\langle   u  p  h' \rangle
         \end{equation}
By \eqref{DD.7a}, $\dis{|\langle u  m' \rangle-\int_{I'} p^{-1}um'|\le c\eps^2}$, having used that
$|m'|$ is bounded, first inequality in \eqref{DD.3}.  Since $m' =p J^{\rm neum}*m' +p  h'$,
using the second inequality in \eqref{DD.3a},
     \begin{equation*}
|  m'(x) -p (J^{\rm neum})'*m(x) |\le  \sup_{y\in I'}| p h' | \le c\eps,\quad x\in I'
         \end{equation*}
Then, by  the second inequality in \eqref{DD.3},
     \begin{equation*}
|  m'(x) -p J^{\rm neum}*\bar m'(x) |= |  m'(x) -p (J^{\rm neum})'*\bar m(x) |\le  c\eps\log\eps^{-1},\quad x\in I'
         \end{equation*}
 and by \eqref{DD.7} and \eqref{DD.7a},
      \begin{equation}
    \label{eD.6-1}
| \langle u  m' \rangle - \langle \tilde m' \bar m'\rangle_{\infty}| \le c\eps\log\eps^{-1}
         \end{equation}
 Analogous estimates hold for $\langle   u  p  h' \rangle$ and  we get
       \begin{equation}
    \label{eD.6-2}
|\la -[1- C_{\eqref{DD.5}}\eps] | \le \frac{c_{\eqref{DD.5}} }2 (\eps \log\eps^{-1})^2
         \end{equation}
 To conclude the proof of the Proposition we need to extend the previous bounds
 to $(\hat h,\hat m)$ in a $\delta$-ball around $(h,m)$.  By \eqref{DD.8.1}
 \begin{equation}
    \label{eD.6-3}
\frac{\hat \la}{\la} \ge \frac{\langle u^2 \rangle}{ \langle u^2 \rangle_{\hat h,\hat m} } \ge c\delta
    \end{equation}
The analogous bound can be proved for $\la/\hat\la$ and
\eqref{DD.5} follows if $\delta$ is small enough.  The proof of Proposition \ref{propDD.1}
is complete.  \qed

\vskip1cm

\noindent
The rest of the spectrum is separated from $\la_{h,m}$
by a  spectral gap, see  \cite{DOP}.

\vskip.5cm

  \begin{prop}
  \label{propDD.2}
There are $ c_{\eqref{DD.28}} $,  $a_{\eqref{DD.28}} >0$, $ c_{\eqref{DD.29}} $ and  $a_{\eqref{DD.29}} >0$
so that for all $\eps$ small
enough  the following holds. For any $(h',m')\in \mathcal A$ there is $\delta=\delta(\eps)$ so that
for all $(h,m)$ in a $\delta$-ball around $(h',m')$,
for
all bounded $\psi$
      \begin{equation}
    \label{DD.28}
\|A^n_{h,m}  \tilde \psi\| \le c_{\eqref{DD.28}} e^{-a_{\eqref{DD.28}} n} \|
\psi\|,\quad \tilde \psi = \psi- \frac{\langle \psi u_{h,m} \rangle_{h,m} }{
\langle u_{h,m}^2\rangle_{h,m} }\,u_{h,m}
         \end{equation}
      \begin{equation}
    \label{DD.29}
|L_{h,m}^{-1}  \tilde \psi(x)| \le c_{\eqref{DD.29}}\int e^{- a_{\eqref{DD.29}}|x-y|}
|\tilde\psi(y)|\, dy
         \end{equation}

  \end{prop}

\vskip.5cm

\centerline{ \em The operator $A^*$ and its spectral properties.}
  \nopagebreak
  \vskip.1cm \noindent
We conclude this appendix with a simple extension of the
previous results which will allow us to complete
the proof of Theorem \ref{thme2.1} and
of  Lemma \ref{lemma4.1}. Let
$(h^*,m^*)$ be the solution of the antisymmetric problem in $\eps^{-1}[-1,\ell^*]$,
with $x_0$ the middle point in $[-1,\ell^*]$.
We denote by $A^*$ the operator $p^*J^{{\rm neum},*}*$
acting on $L^{\infty}(\eps^{-1}[-1,\ell^*])$ with $p^*=p_{h^*,m^*}$ and
kernel $J^{{\rm neum},*}(x,y)$ (defined with Neumann conditions  on $\eps^{-1}[-1,\ell^*]$).
We denote by  $\langle \cdot\rangle_{*}$ the integral over
$\eps^{-1}[-1,\ell^*]$ w.r.t.\ the
measure $(p^*)^{-1}dx$.  We first observe that the pair
$(h^*,m^*)$ satisfies the same properties (with same parameters)
as the pair $(h_\eps,m_\eps)$  (recall that  $m_\eps$  is the restriction of $m^*$ to
$\eps^{-1}[-1,1]$ and that $h_\eps$ is the restriction of $h^*$ except for the
additive term $R_\eps$).  It then follows that $\la^*$ and $u^*$ satisfy the
same properties as $\la_{h,m}$ and $u_{h,m}$ stated in Proposition \ref{propDD.1}
(without loss of generality we may suppose with same coefficients).  Also
Proposition \ref{propDD.2} remains valid, indeed
its validity is quite general as discussed in Section 8.3 of \cite{presutti}.

\vskip.5cm

{\bf Conclusion of the proof of  Theorem \ref{thme2.1}}. In order
to keep the
notation used so far we  replace  the original interval  $\eps^{-1}[-\ell,\ell]$ in
Theorem \ref{thme2.1} by the interval $\eps^{-1}[-1,\ell^*]$ and denote the solution $(h_\eps,m_\eps)$
of  Theorem \ref{thme2.1} by $(h^*,m^*)$.  Reminding that it  only remains to prove
that $m^*(x)$ is an increasing function of $x$ (we are supposing $j<0$),
we shorthand $\psi = \frac{dm^*}{dx}$ and
shall prove that  $\psi(x)$ is strictly positive at all $x$.  We have
   \begin{equation}
    \label{H.0000}
 \psi= L^{-1}\Big(-p^* \frac{dh^*}{dx}\Big)=  L^{-1} (\eps j)
    \end{equation}
where
$L=A^*-1$.  The positivity of $\psi$ then follows from
   \begin{equation}
    \label{H.0001}
 L^{-1} (\eps j) = \sum_{n=0}^\infty (A^*)^n (-\eps j)
    \end{equation}
once we prove that the series  converges (as all its elements are positive).
Convergence follows because there are $a=a(\eps)$ and $c=c(\eps)$  positive such that for all
$n$,
   \begin{equation}
    \label{H.0002}
\| (A^*)^n \| \le c e^{-an}
    \end{equation}
which would be easy if this was the $L^2$ norm as we know that $\la^*$ is the maximal eigenvalue
and $\la^*<1-c\eps$.

\vskip.1cm

\noindent $\bullet$  Proof of \eqref{H.0002}.  With  $\la^*$ and $u^*$  the maximal
eigenvalue and eigenvector of $A^*$, $u^*$ normalized, $\langle(u^*)^2\rangle_*=1$, we have
   \begin{equation}
    \label{H.0003}
(A^*)^n \psi = (\la^*)^n \langle u^* \psi\rangle_* u^* + (A^*)^n \tilde \psi,\quad
\tilde\psi = \psi-\langle u^* \psi\rangle_* u^*
    \end{equation}
We have  $\la^*<1-C\eps$, $C>0$, (by \eqref{DD.5}),
we bound $u^*$ using \eqref{DD.7a}, then by  \eqref{DD.28}
   \begin{equation}
    \label{H.0004}
\|(A^*)^n \psi \|\le c (\la^*)^n \| \psi\| + c_{\eqref{DD.28}} e^{-a_{\eqref{DD.28}} n} \|
\psi\|
    \end{equation}
hence \eqref{H.0002}.  \qed

\vskip1cm

{\bf Conclusion of the proof of Lemma \ref{lemma4.1}}. It   only remains to prove the second inequality in
\eqref{4z.3}. With $\psi = \frac{dm^*}{dx}$, by \eqref{H.0000}, and using the previous notation
   \begin{equation}
    \label{H.00}
 \psi= \Big([\la^*-1]^{-1}\eps j \int_{-\eps^{-1}}^{\eps^{-1}\ell^*} u^* \Big) u^* + L^{-1} \phi,\quad \phi= (\eps j)-(\eps j \int u^* ) u^*
    \end{equation}
$|[\la^*-1]^{-1}\eps j|\le c$ by \eqref{DD.5} and by  \eqref{DD.7a}:
   \begin{equation*}
 \int u^* \le c, \qquad \sup_{|x-\eps^{-1}x_0|\ge
r_{\eqref{4z.3}}\log\eps^{-1}}u^*(x) \le
c_{\eqref{DD.7a}} e^{-a_{\eqref{DD.4}}r_{\eqref{4z.3}}\log\eps^{-1}}
    \end{equation*}
  By choosing $a_{\eqref{DD.4}}r_{\eqref{4z.3}}>1$
the first term on the r.h.s.\ of \eqref{H.00} is bounded by $c\eps$
when  $|x-\eps^{-1}x_0|\ge
r_{\eqref{4z.3}}\log\eps^{-1}$.  The last term is bounded using \eqref{DD.28} by
   \begin{equation*}
\le c_{\eqref{DD.28}} \sum_{n=0}^\infty e^{-a_{\eqref{DD.28}} n}\|\phi\| \le c'\eps
    \end{equation*}
because $\|\phi\| \le c\eps$.   Lemma \ref{lemma4.1} is   proved.  \qed

\vskip2cm

\section{\hskip.2cm An auxiliary dynamics}

  \label{appE}

We return in this appendix to the analysis
of the auxiliary dynamics introduced in
Appendix \ref{appA}.  We shall study the case
where  initially $(h_0,m_0)\in \mathcal A$
and prove a local existence and uniqueness theorem under
suitable assumptions on $h(t)$.  We would like to
work in  $\mathcal A$ but  $\mathcal A$ itself
is not  nice
in the $L^\infty$ topology  we are using
as it involves  derivatives. For this reason  we introduced the $\delta$-balls
of  $\mathcal A$ in the previous appendix which will play an important role here as well.
Our first result
is a straight consequence of Proposition
\ref{propDD.1} and  Proposition
\ref{propDD.2} and its proof is omitted:

\vskip.5cm

  \begin{prop}
  \label{propE.1}
There is $c>0$ and for any
$\eps>0$ small enough
there is  $\delta=\delta(\eps)>0$ not larger than the parameter $\delta$ in Proposition \ref{propDD.1}
so that for any  $(h_0,m_0)\in \mathcal A$
and any  $(h',m')$ and $(h'',m'')$ in the $\delta$-ball of $(h_0,m_0)$
   \begin{equation}
    \label{E.2}
    \|L_{h',m'}^{-1}\|\le c\eps^{-1},\quad
\|L_{h',m'}^{-1} - L_{h'',m''}^{-1}\| \le c \eps^{-2} \big(\|h'-h''\|+\|m'-m''\|\big)
    \end{equation}

  \end{prop}

\vskip.5cm

  \begin{prop}
  \label{propE.2}
Let $\delta$ and $c$ be
as in Proposition \ref{propE.1} and let $C$ be any positive number.  Then
for any $\eps>0$ small enough  there
is $T \in (0, \frac \delta {2C})$ so that the following holds. For any $(h_0,m_0)\in \mathcal A$,
and any $h(t)$, $t\in [0,T]$, such that
$h(0)=h_0$ and
$\|\frac{dh(t)}{dt}\| \le C$
there is $m(t)$, $t\in [0,T]$, such that:
   \begin{equation}
    \label{E.3}
\frac{dm}{dt}=  L_t^{-1}\big(- p_t
\frac{dh}{dt}\big),\;\; m(0)=m_0,\qquad L_t=L_{h(t),m(t)}, \, p_t=p_{h(t),m(t)}
    \end{equation}
($L_t=L_{h(t),m(t)}$, $p_t=p_{h(t),m(t)}$), $\|m(\cdot)-m_0\|\le \frac \delta 2$ and  $m(\cdot)=\tanh\{\beta J^{\rm neum}*m(\cdot)+\beta h(\cdot)\}$.  Finally $m(\cdot)$ is the unique solution of \eqref{E.3} in
$\|m(\cdot)-m_0\|\le \frac \delta 2$.

  \end{prop}

\vskip.5cm

\noindent {\bf Proof.} $T$ is determined by the following three conditions:
   \begin{equation}
    \label{E.4}
T < \frac\delta {2C},\;\;c\eps^{-1} \beta CT < \frac\delta {2},\;\;\Big(c\eps^{-1} 2\beta^2+ \beta c \eps^{-2}\Big)T<1
    \end{equation}
The first one ensures that $\|h(\cdot)-h_0\|< \frac \delta 2$
(because $\|\frac{dh(t)}{dt}\| \le C$); the second one (obtained by bounding $L_t^{-1}\big(- p_t
\frac{dh}{dt}\big)$ via Proposition \ref{propE.1}) will imply that
$\|m(\cdot)-m_0\|< \frac \delta 2$, so that $(h(t),m(t))$ is always in the $\delta$-ball of $(h_0,m_0)$
and  Proposition \ref{propE.1} can be applied. The third condition will imply that the integral
version of \eqref{E.3} gives rise to a contraction.

\noindent Let $\dis{\mathcal X:=\big\{ m\in C\big([0,T], L^\infty(\eps^{-1}[-1,1]; [-1,1])\big):
m(0)=m_0, \|m(\cdot)-m_0\| \le \frac\delta 2\big\}}$ and for $m\in \mathcal X$ let
   \begin{equation}
    \label{E.5}
\psi(m)(t)=m_0+ \int_0^t L^{-1}_s \big(-p_s \frac{dh(s)}{ds}\big)
    \end{equation}
By \eqref{E.2} and  the second
inequality in \eqref{E.4}, $\dis{\|\psi(m)(t)- m_0\| \le c\eps^{-1} \beta C t <\frac\delta 2}$.  Thus $\psi$ maps $\mathcal X$ into itself.  By \eqref{E.2} and the
third inequality in \eqref{E.4} $\psi$ is a contraction with sup norm in $x$ and $t$.  Therefore
there is a fixed point $m\in \mathcal X$: $m=\psi(m)$ and since $\psi$ maps $\mathcal X$ into
functions which are differentiable in $t$ with bounded derivative, $m$ is a solution of
\eqref{E.3}.  By \eqref{E.3}
   \begin{equation*}
 \frac{d}{dt}\Big(   m(t) - \tanh\{\beta J^{\rm neum}*m(t)+\beta h(t)\}\Big) =0
    \end{equation*}
so that  $ m(t) - \tanh\{\beta J^{\rm neum}*m(t)+\beta h(t)\} =
m_0 - \tanh\{\beta J^{\rm neum}*m_0+\beta h_0\}=0$.  Finally, if $m$ solves \eqref{E.3}
and $\|m(\cdot)-m_0\|\le \frac \delta 2$, then $m\in \mathcal X$ and $\psi(m)=m$.  Since
$\psi$ is a contraction $m$ is unique.   \qed

\vskip2cm

\section{\hskip.2cm Properties of the sets $\mathcal A$ and $\mathcal G$}

  \label{appF}

The intervals $I$ and $I'$ which appear frequently in the sequel
have been defined in \eqref{DD.00}.

  \vskip.5cm

  \centerline{\em Fixing the parameters in the set
  $\mathcal G$}
 \nopagebreak \vskip.1cm \noindent
 The coefficient
$a_{\eqref{e4.8a}}$ is a  positive number strictly smaller
than all the parameters in Appendix \ref{appDD} involved with
exponential decay, in particular we
require $a_{\eqref{e4.8a}} < \min\{a _{\eqref{DD.4}}(1-x_0),a_{\eqref{DD.29}}\frac{1+x_0}{1-x_0}\}$.  The other parameter $b_{\eqref{e4.8az}}$ is fixed so that:
   \begin{equation}
    \label{FF.0}
b_{\eqref{e4.8az}}<1\;\;\text{and such that}\;\;\; \{\frac{3 \beta  c_{\eqref{DD.29}}}{a_{\eqref{DD.29}}-a_{\eqref{e4.8a}} \frac{1-x_0}{1+x_0}}\}\, b_{\eqref{e4.8az}}\,
< \frac{e^{-2a_{\eqref{DD.4}}}-e^{-4a_{\eqref{DD.4}}}}{2\beta}
  \end{equation}

\vskip.5cm

  \begin{prop}
  \label{propF.1}
For all $\eps>0$ small enough if $h\in \mathcal G$ there is $m$ such that
$m=\tanh\{\beta J^{\rm neum}*m+\beta h\}$ and $(h,m)\in \mathcal A$.

  \end{prop}

\vskip.5cm

\noindent {\bf Proof.}  Given $h\in \mathcal G$ and $t\in [0,1]$ we define $h(t):= t h +(1-t)h_\eps$
observing that
$(h(0),m(0)):=
(h_\eps,m_\eps)\in \mathcal A$ by the definition of $\mathcal A$.  Let $S$ be the sup of
all $s\le 1$ such that there exists $m(t)$, $t\in [0,s]$, which solves \eqref{E.3}
in $[0,s]$ starting from $m(0)=m_\eps$
and such that for all such $t$, $(h(t),m(t))$ is in the $\delta$-ball of $\mathcal A$ with $\delta$ as
Proposition  \ref{propDD.1}.  We shall prove that $S=1$ and that for all $t\le 1$
$(h(t),m(t))\in\mathcal A$ thus proving the Proposition.

\noindent Since $\|\frac{dh(t)}{dt}\|=\|h-h_\eps\| \le  b_{\eqref{e4.8az}}$ (because $h\in \mathcal G$)
we can apply Proposition \ref{propE.2} with $C=b_{\eqref{e4.8az}}$ and $(h_0,m_0)=
(h_\eps,m_\eps)\in \mathcal A$. As a consequence there is
$T=T(\eps)>0$ so that $m(t)= \tanh\{\beta J^{\rm neum}*m(t)+ \beta h(t)\}$, $t\in [0,T]$,
and $\delta=\delta(\eps)$
so that $\|h(t)-h_\eps\|\le \delta/2$, $\|m(t)-m_\eps\| \le\delta$/2, $t\in [0,T]$.
Since $\delta$ is not larger than the
parameter $\delta$ of Proposition  \ref{propDD.1} (see Proposition  \ref{propE.1}) we then conclude
that
$S\ge T$. By the definition of $S$ the bounds in
Proposition  \ref{propDD.1} hold for $(h(t),m(t))$ at any  $t\in [0,S]$ and
it is now just a matter of computations to check
that $(h(t),m(t))\in \mathcal A$ for all such $t$.  We start by proving that $h(t)$
satisfies the conditions in \eqref{DD.3a}.
   \begin{equation*}
\|h(t)\| \le \|h_\eps\|+\|h-h_\eps\|\le  \frac {C_{\eqref{DD.3a}}}2 + b_{\eqref{e4.8az}} \le
C_{\eqref{DD.3a}},\quad \text{(as $b_{\eqref{e4.8az}}<1<\frac {C_{\eqref{DD.3a}}}2$)}
    \end{equation*}
   \begin{equation*}
\|\frac {dh(t)}{dx}\| \le \|\frac {dh_\eps}{dx}\|+\|\frac {d(h-h_\eps)}{dx}\|\le  \frac {C_{\eqref{DD.3a}}}2 + \eps \le
C_{\eqref{DD.3a}}
    \end{equation*}
for $\eps$ small enough.  Finally  in $I$ (defined in  \eqref{DD.00})
    \begin{equation*}
|\frac{dh(t)}{dx}  - \frac{-\eps j}{\bar p_{x_0}}|\le |\frac{dh_\eps}{dx} - \frac{-\eps j}{\bar p_{x_0}}| + t |\frac{d(h-h_\eps)}{dx}| \le
\frac{ c_{\eqref{DD.3a}}}2 \eps^2\log\eps^{-1} +\eps^2 \le  c_{\eqref{DD.3a}}\eps^2\log\eps^{-1}
    \end{equation*}
(for $\eps$ small enough)
so that also the last condition in \eqref{DD.3a} is satisfied.

We shall next prove that $m(t)$ satisfies the conditions required in $\mathcal A$.
We write
$f(t):=- p_{t} [h-h_\eps]$; $\la_t$, $u(t)$ for the maximal eigenvalue and eigenvector of $A_t$;
$\langle \cdot \rangle_t$ for the integral of the measure $p_t^{-1}dx$ on $\eps^{-1}[-1,1]$; $\tilde f(t):=f(t)-
 \langle
u(t)f(t)\rangle_t u(t)$. By \eqref{E.3}
   \begin{equation}
    \label{FF.1}
\frac{dm(t)}{dt} = L_{t}^{-1} f(t) = \la_t^{-1}  \langle
u(t)f(t)\rangle_t u(t) +  L_{t}^{-1} \tilde f(t),\quad  \langle
u(t)^2\rangle_t=1
    \end{equation}
We bound $|f(t)| \le \beta N(h-h_\eps) E_\eps(x)^{-1}$ (using that $h\in \mathcal G$ and $p_t\le \beta$),
$u(t)\le  c_{\eqref{DD.7a}}e^{-a _{\eqref{DD.4}}|x-\eps^{-1}x_0|}$ and get
   \begin{eqnarray}
    \label{FF.1.1}
&& | \langle
u(t)f(t)\rangle_t |\le c  N(h-h_\eps) e^{-a_{\eqref{e4.8a}}(1-x_0)\eps^{-1}} \nn\\&&
|\tilde f(t)| \le   N(h-h_\eps) \Big(ce^{-a_{\eqref{e4.8a}}(1-x_0)\eps^{-1}} c_{\eqref{DD.7a}}e^{-a _{\eqref{DD.4}}|x-\eps^{-1}x_0|}
+ \beta  E_\eps^{-1}\Big)
    \end{eqnarray}
By \eqref{DD.29} $\dis{| L_{t}^{-1} \tilde f(t)|(x)\le  c_{\eqref{DD.29}}\int e^{- a_{\eqref{DD.29}}|x-y|}
 |\tilde f(t)|\, dy}$ and
   \begin{equation*}
\int e^{- a_{\eqref{DD.29}}|x-y|} E_\eps(y)^{-1} \le \frac{2 E_\eps(x)^{-1}}{a_{\eqref{DD.29}}-a_{\eqref{e4.8a}}}
    \end{equation*}
so that, by
\eqref{FF.1} and  \eqref{FF.1.1} and since $\la_t \le c\eps^{-1}$
   \begin{eqnarray}
    \nn
|m(t)-m_\eps|\le  N(h-h_\eps) \Big(&& c'  \eps^{-1}e^{-a_{\eqref{e4.8a}}(1-x_0)\eps^{-1}} c_{\eqref{DD.7a}}e^{-a _{\eqref{DD.4}}|x-\eps^{-1}x_0|} \\&&+ \frac{2 \beta  c_{\eqref{DD.29}} E_\eps(x)^{-1}}{a_{\eqref{DD.29}}-a^-_{\eqref{e4.8a}}}\Big)
    \label{FF.2}
    \end{eqnarray}
which (recalling that $N(h-h_\eps) \le b_{\eqref{e4.8az}}$) proves that for $\eps$ small enough,
       \begin{eqnarray}
           \label{FF.2.11.00}
&&  \|m(t)-m_\eps\| \le
\kappa:= \frac{3 \beta  c_{\eqref{DD.29}}b_{\eqref{e4.8az}}}{a_{\eqref{DD.29}}-a^-_{\eqref{e4.8a}}}\\
    \label{FF.2.11}
&&
N(m(t)-m_\eps)\le \Big( c'  \eps^{-1}c_{\eqref{DD.7a}} + \frac{2 \beta  c_{\eqref{DD.29}} }{a_{\eqref{DD.29}}-a^-_{\eqref{e4.8a}}}\Big) N(h-h_\eps)
    \end{eqnarray}
By \eqref{FF.2.11.00}, $\dis{p_t \le \beta \big(1-(|m_\eps|-\kappa)^2\big) < e^{-4 a_{\eqref{DD.4}}} + 2\beta   \kappa}$ in
$\{|x-\eps^{-1}x_0|\ge r_{\eqref{DD.4}}\}$ and therefore, by \eqref{FF.0},  $m(t)$  satisfies  \eqref{DD.4}
for $t\in [0, T]$.  To prove the second condition in \eqref{DD.3} we write for $x\in I$,
      \begin{eqnarray*}
  |m(x,t)-\bar m_{x_0}(x)| &\le & |m(x,t)-m_\eps(x)| + |m_\eps(x)-\bar m_{x_0}(x)|
\\
&\le & N(m(t)-m_\eps) E_\eps(x)^{-1} +  c_{\eqref{zD.2}} \eps\log\eps^{-1}
\\
&\le & c \eps^{-1} E_\eps(x)^{-1} +  c_{\eqref{zD.2}} \eps\log\eps^{-1} \le 2 c_{\eqref{zD.2}} \eps\log\eps^{-1}
    \end{eqnarray*}
for $\eps$ small enough (because $E_\eps(x)^{-1} \le e^{-a^-_{\eqref{e4.8a}} [\eps^{-1}(1-x_0) -2C_{\eqref{DD.2}} \log\eps^{-1}]}$).  The second inequality in \eqref{DD.3} then follows recalling that
$c'_{\eqref{DD.3}} > 2c_{\eqref{zD.2}}$. To prove the first inequality in \eqref{DD.3}
we take
the $x$-derivative of the equality  $m(t)= \tanh\{\beta J^{\rm neum}*m(t)+ \beta h(t)\}$:
   \begin{eqnarray}
    \nn
&&\frac{dm(t)}{dx} =  \la_t^{-1}  \langle
u(t)g(t)\rangle_t\, u(t) +  L_{t}^{-1} \tilde g(t)\\&& g(t):= \{- p_{t}\frac{d [h-h_\eps]}{dx}\},\;
\tilde g(t)=g(t)- \langle
u(t)g(t)\rangle_t\, u(t)
\label{FF.3}
    \end{eqnarray}
By \eqref{e4.8az}, $\|g(t)\|\le \beta\eps$ and an argument similar to the previous one shows that
$\|\frac{dm(t)}{dx}\| \le c\eps$, so that
also the first condition in \eqref{DD.3}
is satisfied.

In conclusion we have   proved so far
that for all $\eps$ small enough, $(h(t),m(t))
\in \mathcal A$ for all $t\in [0,S]$.  Suppose by contradiction that $S<1$,
write $S'= \min\{1,S+T\}$, then
since $(h(S),m(S))\in \mathcal A$ by Proposition \ref{propE.2}
there is $m(t)$, $t\in [S,S']$, which solves
\eqref{E.3}
in $[S,S']$ starting from $m(S)$
and such that for all such $t$, $(h(t),m(t))$ is in the $\delta$-ball of
$(h(S),m(S))$, hence, a fortiori, in the  $\delta$-ball of $\mathcal A$ with $\delta$
as in  Proposition  \ref{propDD.1}.  This contradicts the maximality of $S$ hence
$S=1$.   \qed

\vskip1cm

  \begin{prop}
  \label{propF.2}
There are $a_{\eqref{FF.4}}>0$, $r_{\eqref{FF.4}}>0$, $c$, $c'$ and $a_{\eqref{FF.5}}>0$ so that
for all $\eps$ small enough the following holds. Let $h\in \mathcal G$ and $(h,m)\in \mathcal A$ (existence of $m$
follows from Proposition \ref{propF.1}),
then
   \begin{equation}
    \label{FF.4}
\sup_{|x-\eps^{-1}x_0|\le 2r_{\eqref{FF.4}} \eps^{-1}} |m(x)-m_\eps(x)| \le e^{-a_{\eqref{FF.4}}\eps^{-1}}
    \end{equation}
   \begin{equation}
    \label{FF.5}
|\la_{h,m} -\la_\eps |  \le c e^{-a_{\eqref{FF.4}}\eps^{-1}},\quad
\|u_{h,m} -u_\eps \|
\le
c'e^{-a_{\eqref{FF.5}}\eps^{-1}}
    \end{equation}

  \end{prop}

\vskip.5cm

\noindent {\bf Proof.}   \eqref{FF.4} follows from \eqref{FF.2.11}, in
the sequel it is convenient to have
$a_{\eqref{FF.4}}$ small, in particular $a_{\eqref{FF.4}} < a_{\eqref{DD.7a}}$.
Let  $\la$ and $u$ be the
maximal eigenvalue and  eigenvector of $A:=A_{h,m}$, $u>0$ normalized so that
$\langle u^2\rangle =1$ ($\langle \cdot \rangle:=\langle \cdot\rangle_{h,m}$).
and $\la_\eps$, $u_\eps$ the maximal
eigenvalue and  eigenvector of
$A_\eps:=A_{h_\eps,m_\eps}$ with $u_\eps>0$ normalized so that
$\langle u_\eps^2\rangle_\eps =1$ ($\langle \cdot \rangle_\eps
:=\langle \cdot\rangle_{h_\eps,m_\eps}$).
Since $(h_\eps,m_\eps)$
and $(h,m)$ are both in $\mathcal A$ we can use the bounds established in Proposition \ref{propDD.1}
and \ref{propDD.2}
for $A$ and $A_\eps$.

\noindent We then have
   \begin{equation}
    \label{FF.6}
\frac{ \la }{\la_\eps} \ge \frac{\langle u_\eps^2\rangle_\eps}{\langle u_\eps^2\rangle}
= 1 - \frac{\langle u_\eps^2(1- \frac{p}{p_\eps})\rangle}{\langle u_\eps^2\rangle}\ge 1 - c
e^{-a_{\eqref{FF.4}}\eps^{-1}}
    \end{equation}
the first inequality following
from \eqref{DD.8.1}.  To prove the last one we recall that $p_\eps=p_{h^*,m^*}\ge c_{\eqref{e4.0}}$,
$p\equiv p_{h,m} \ge C_{\eqref{DD.5z}}$. In $\{x:|x-\eps^{-1}x_0|\le 2r_{\eqref{FF.4}} \eps^{-1}\}$
we use \eqref{FF.4} to get
    \begin{equation}
    \label{aFF.6}
 \sup_{|x-\eps^{-1}x_0|\le 2r_{\eqref{FF.4}} \eps^{-1}}
|1- \frac{p}{p_\eps}|   \le c e^{-a_{\eqref{FF.4}}\eps^{-1}}
    \end{equation}
In  $\{x:|x-\eps^{-1}x_0|> 2r_{\eqref{FF.4}} \eps^{-1}\}$ we bound $\dis{|1- \frac{p}{p_\eps}|   \le \frac{2\beta}{
c_{\eqref{e4.0}} C_{\eqref{DD.5z}}}}$ and $u_\eps$
using \eqref{DD.7a}.
Same argument is used to bound
from below $\dis{\frac{\la_\eps}{\la }}$ and the first inequality in \eqref{FF.5}
follows because $\la$ and $\la_\eps$ are both close to 1 by $c\eps$.

In order to compute the sup in the second inequality in
\eqref{FF.5} we consider  first
$|x-\eps^{-1}x_0|> r_{\eqref{FF.4}} \eps^{-1}$.  In such a case both $u$ and $u_\eps$ are smaller
than $ c_{\eqref{DD.7a}} e^{-a _{\eqref{DD.4}}\eps^{-1}r_{\eqref{FF.4}}}$ hence
their difference is bounded by $c'e^{-a_{\eqref{FF.5}}\eps^{-1}}$,
provided $a_{\eqref{FF.5}}<a _{\eqref{DD.4}}
r_{\eqref{FF.4}}$.
We next take $|x-\eps^{-1}x_0|\le r_{\eqref{FF.4}} \eps^{-1}$.
Analogously to \eqref{DD.15} and with $y_0\equiv x$,
  \begin{equation}
    \label{FF.7}
\la ^N u (x)=  \int  u (y_N)
 \prod_{k=1}^N \{ A_\eps(y_{k-1},y_k) \frac{p (y_{k-1})}
 {p_\eps (y_{k-1})}\}  dy_1\cdots dy_N
    \end{equation}
We choose $N=b \eps^{-1}$ with $b>0$ smaller than $r_{\eqref{FF.4}}$.
Then for all $k\le N$,  $|y_k-\eps^{-1}x_0| \le (r_{\eqref{FF.4}}+b) \eps^{-1} \le
2 r_{\eqref{FF.4}} \eps^{-1}$ so that by \eqref{aFF.6}  for all $\eps$ small enough,
$\dis{ u (x)\le \la^{-N} [1 + c  e^{-a_{\eqref{FF.4}}\eps^{-1}} ]^N
A_\eps^N u(x)}$.
We then write
$A_\eps^N u(x) = \la_\eps^N  \langle uu_\eps\rangle_\eps u_\eps
+ A_\eps^N \tilde u(x)$ so that in  $\{|x-\eps^{-1}x_0|\le r_{\eqref{FF.4}} \eps^{-1}\}$
  \begin{eqnarray*}
   u\le  [1 + c  e^{-a_{\eqref{FF.4}}\eps^{-1}} ]^N
   \Big((\frac{ \la_\eps}{\la})^N   \langle uu_\eps\rangle_\eps u_\eps
  + \la^{-N} \|u\| c_{\eqref{DD.28}} e^{-a_{\eqref{DD.28}} N}\Big)
    \end{eqnarray*}
By  \eqref{DD.16b}
  \begin{eqnarray}
    \label{FF.9}
 u\le  [1 + c  e^{-a_{\eqref{FF.4}}\eps^{-1}} ]^N
   \Big((\frac{ \la_\eps}{\la})^N   \langle uu_\eps\rangle_\eps u_\eps
  + \la^{-N}   c  e^{-a_{\eqref{DD.28}} N}\Big)
    \end{eqnarray}
We bound $\dis{  \langle uu_\eps\rangle_\eps  \le  \langle u^2\rangle_\eps^{1/2},\;\;\langle u^2\rangle_\eps=
  1-\langle u^2|1-\frac{p}{p_\eps}|\rangle}$ and use the previous bounds for $|1-\frac{p}{p_\eps}|$
so that $  \langle uu_\eps\rangle_\eps  \le 1 + c e^{-a\eps^{-1}}$
with $a$ and $c$ suitable positive constants.  By \eqref{FF.6}
  \begin{equation*}
(\frac{\la_\eps} { \la })^N\le e^{- N \log\{1- c
e^{-a_{\eqref{FF.4}}\eps^{-1}}\}} \le \exp\{c' \eps^{-1}e^{a_{\eqref{FF.4}}\eps^{-1}}\}
\le 1 + c'' \eps^{-1}e^{a_{\eqref{FF.4}}\eps^{-1}}
    \end{equation*}
Collecting all these bounds and recalling that $\la<1-c\eps$, we get from \eqref{FF.9}
  \begin{eqnarray}
    \label{FF.99}
 u(x)  &\le &\big(1+c \eps^{-1}
   e^{-a_{\eqref{FF.4}}\eps^{-1}}\big)
   \, u_\eps(x)\nn\\&&
  +c e^{ -N\big(\log(1-c\eps)-a_{\eqref{DD.28}} \big)}
    \end{eqnarray}
hence the upper bound for $u$ in \eqref{FF.5}.  The lower bound is proved similarly.  \qed

\vskip1cm

\noindent
Recall that $(h^*,m^*)$ is the solution of the antisymmetric problem in $\eps^{-1}[-1,\ell^*]$,
with $x_0$ the middle point in $[-1,\ell^*]$
and $m_\eps$  the restriction of $m^*$ to
$\eps^{-1}[-1,1]$.  We denote by $\la^*$ and $u^*$ the maximal eigenvalue and
eigenvector of $A_{h^*,m^*}$ and by  $\la_\eps$ and $u_\eps$ those of $A_\eps= A_{h_\eps,m_\eps}$,
writing   $u_\eps$ also for its extension to  $\eps^{-1}[-1,\ell^*]$ with
$u_\eps=0$ outside $\eps^{-1}[-1,1]$.  We suppose  $\langle u_\eps^2\rangle_\eps=
\langle (u^*)^2\rangle_{*}=1$ with the obvious meaning of the symbols.

\vskip.5cm

  \begin{prop}
  \label{propF.3}
For all $\eps$ small enough,
   \begin{equation}
    \label{FF.11}
|\la^* -\la_\eps |  \le c e^{-a _{\eqref{DD.4}}\eps^{-1}(1-x_0)},\quad
\|u^* -u_\eps \|
\le
c_{\eqref{FF.11}} e^{-a_{\eqref{FF.11}}\eps^{-1}(1-x_0)}
    \end{equation}

  \end{prop}

\vskip.5cm

\noindent {\bf Proof.}
Since $p_\eps=p^*$ in $\eps^{-1}[-1,1]$, $\langle u_\eps^2\rangle_{*}=\langle u_\eps^2\rangle_\eps=1$
so that $\dis{ \la^*  \ge \int  u_\eps  J^{{\rm neum},*}* u_\eps }$ and, by \eqref{DD.7a},
       \begin{equation}
    \label{FF.12}
\|(J^{{\rm neum},*}-J^{{\rm neum},\eps})*u_\eps\|  \le   c e^{-a _{\eqref{DD.4}}\eps^{-1}(1-x_0)}
         \end{equation}
where $J^{{\rm neum},\eps}$ and $J^{{\rm neum},*}$ are the kernel with Neumann conditions respectively
in $\eps^{-1}[-1,1]$ and $\eps^{-1}[-1,\ell^*]$. Thus
   \begin{equation}
    \label{aFF.13}
 \la^*  \ge \int  u_\eps J^{{\rm neum},\eps}* u_\eps - c e^{-a _{\eqref{DD.4}}\eps^{-1}(1-x_0)}
 \ge \la_\eps - c e^{-a _{\eqref{DD.4}}\eps^{-1}(1-x_0)}
    \end{equation}
For the reverse inequality we write $\dis{  \la_\eps  \ge \frac{\int
u^*  J^{{\rm neum},\eps} u^* }{\langle (u^*)^2\rangle_\eps}}$, the
integral being extended to $\eps^{-1}[-1,1]$. Using \eqref{FF.12} we
replace the kernel $ J^{{\rm neum},\eps}$ with $ J^{{\rm neum},*}$
and then extend the integral to  $\eps^{-1}[-1,\ell^*]$ bounding
$u^*$ via \eqref{DD.7a} which holds as well for $u^*$ in the whole
$\eps^{-1}[-1,\ell^*]$  (see the paragraph ``The operator $A^*$ and
its spectral properties'' at the end of Appendix \ref{appDD}). In
this way we derive the first inequality in \eqref{FF.11}.

As in the proof of Proposition \ref{propF.2}
we bound $|u^*(x) -u_\eps(x)|\le c'e^{-a_{\eqref{FF.11}}\eps^{-1}}$ when
$|x-\eps^{-1}x_0|> r_{\eqref{FF.4}} \eps^{-1}$  using \eqref{DD.7a} (supposing
 $a_{\eqref{FF.11}}<a _{\eqref{DD.4}}
r_{\eqref{FF.4}}$).
When $|x- \eps^{-1}x_0|  \le r_{\eqref{FF.4}} \eps^{-1}$ we write
   \begin{equation}
    \label{FF.13}
 u^*(x)= (\la^*)^{-N}(A^*)^N   u^* (x)=(\la^*)^{-N} A_\eps^N   u^* (x)
    \end{equation}
provided $(x_0+r_{\eqref{FF.4}}) \eps^{-1} +N \le \eps^{-1}$, which is satisfied if $N=a\eps^{-1}$
with $a>0$ small enough. Hence
   \begin{equation}
    \label{FF.14}
 u^*(x)= (\frac{\la_\eps}{\la^*})^{N}\langle u^* u_\eps\rangle_\eps u_\eps (x)+ A_\eps^N \tilde u^*
    \end{equation}
   \begin{equation}
    \label{FF.15}
 |u^*(x)- (\frac{\la_\eps}{\la^*})^{N}\langle u^* u_\eps\rangle_\eps u_\eps (x)| \le c
 e^{-a_{\eqref{DD.28}} N}
    \end{equation}
By \eqref{FF.15} and since by   \eqref{DD.7a}  $\langle u^*\rangle_\eps\le c$ and
$|\langle (u^*)^2\rangle_\eps-1|\le c e^{-a_{\eqref{DD.4}}\eps^{-1}(1-x_0)}$
   \begin{equation}
    \label{FF.16}
 |1- (\frac{\la_\eps}{\la^*})^{N}\langle u^* u_\eps\rangle_\eps ^2| \le c
 e^{-a_{\eqref{DD.28}} N} +c e^{-a_{\eqref{DD.4}}\eps^{-1}(1-x_0)}
    \end{equation}
so that the second inequality in \eqref{FF.11} follows from the first one.   \qed

\vskip.5cm

\noindent
As a corollary of Proposition \ref{propF.2}  and  Proposition \ref{propF.3} we have:

\vskip.5cm

  \begin{coro}
  \label{coroF.1}
In the same context of Proposition \ref{propF.2},
   \begin{equation}
    \label{FF.17}
|\la^* -\la_{h,m} |  \le c e^{-a _{\eqref{DD.4}}\eps^{-1}(1-x_0)},\quad
\|u^* -u_{h,m}\|
\le
c_{\eqref{FF.11}} e^{-a_{\eqref{FF.11}}\eps^{-1}(1-x_0)}
    \end{equation}

  \end{coro}

\vskip2cm

\section{\hskip.2cm Convergence of the iterative scheme}

  \label{appH}

By \eqref{e4.6} with $n=-1$ we have for $x\in \eps^{-1}[-1,1]$,
      \begin{equation}
    \label{eG.1}
\hat h_{0}(x) = -\eps j\int_{\eps^{-1}x_0}^x \chi(m_\eps(y))^{-1}
= h^*(x)
    \end{equation}
because $m_\eps=m^*$ on $\eps^{-1}[-1,1]$ and $(h^*,m^*)$ is a solution of \eqref{e2.22}
in $\eps^{-1}[-1,\ell^*]$.
Thus  by \eqref{e4.6}
      \begin{equation}
   \label{e4.6.1}
h_{0}(x) = h^*(x) - \frac{ \int  h^* u^*}
{\int u^*}
    \end{equation}
where the integrals are extended to $\eps^{-1}[-1,1]$.  Then, recalling \eqref{e4.1},
      \begin{equation}
   \label{e4.6.2}
h_{0}(x)-h_{\eps}(x) = -R_\eps(x) - \frac{ \int  h^* u^*}
{\int u^*}
    \end{equation}

\vskip1cm

  \begin{prop}
  \label{propH.1}
For all $\eps$ small enough  $h_0\in \mathcal G$ and
   \begin{equation}
    \label{H.1}
    N(h_0-h_\eps) \le c_{\eqref{H.1}} \eps
    \end{equation}

  \end{prop}

\noindent {\bf Proof.} $\dis{\int_
  {\eps^{-1}(2x_0-1)}^{\eps^{-1}} h^*u^*  =0}$ because
$h^*$ is antisymmetric and  $u^* $ symmetric around the middle point
$\eps^{-1}x_0$ of the interval
$\eps^{-1}[-1,\ell^*]$ ($u^*$ is symmetric because the eigenvalue $\la^*$ is simple
and $A^*$ symmetric).  Since the estimates in Proposition
\ref{propDD.1} apply to   $u^*$ as well (see
the paragraph {\em The operator $A^*$ and its spectral properties} at the end
of Appendix \ref{appDD}),  by
\eqref{DD.7a} and since $\|h^*\| \le c$ we get
   \begin{equation}
    \label{H.1.1}
   \int_{-\eps^{-1}}^{\eps^{-1}} u^*h^*  =
  \int_{-\eps^{-1}}
  ^{\eps^{-1}(2x_0-1)}  h^*u^* \le ce^{-a _{\eqref{DD.4}}\eps^{-1}(1-x_0)}
    \end{equation}
Recalling that $c_{\eqref{e4.0}}$ in \eqref{e4.0} is strictly positive uniformly in $\eps$,
we shall next prove that
       \begin{equation}
    \label{H.1.111}
   \int_{-\eps^{-1}}^{\eps^{-1}} u^* \ge \frac{c_{\eqref{e4.0}}}{ c_{\eqref{DD.7a}}}
    \end{equation}
By \eqref{DD.7a}
$ u^*\le c_{\eqref{DD.7a}}$, \eqref{H.1.111} then follows from \eqref{e4.0}:
       \begin{equation*}
1= \langle (u^*)^2\rangle_*=    \int \frac{(u^*)^2}{p^*}\le   \int \frac{u^* c_{\eqref{DD.7a}}}{c_{\eqref{e4.0}}}
= \{ \frac{c_{\eqref{DD.7a}}}{c_{\eqref{e4.0}}}\} \int u^*
    \end{equation*}
Thus, recalling \eqref{e4.8a} and that $a _{\eqref{DD.4}}(1-x_0) > a_{\eqref{e4.8a}}$, see
the paragraph {\em Fixing the parameters in
the set $\mathcal G$} in
Appendix \ref{appF},
   \begin{equation}
    \label{H.1.2}
N\Big( \frac{ \int  h^* u^*}
{\int u^*}\Big) =\sup_{|x|\le \eps^{-1}} E_\eps(x)\frac{ \int  h^* u^*}
{\int u^*} \le  c  e^{-a_{\eqref{H.1.2}} \eps^{-1}}
    \end{equation}
with   $0<a_{\eqref{H.1.2}} < a _{\eqref{DD.4}}(1-x_0) - a_{\eqref{e4.8a}}$.
By   Lemma
\ref{lemma4.1}, $N(R_\eps) \le c\eps$ which together with \eqref{H.1.2} proves
\eqref{H.1}.  Before proving that $h_0\in \mathcal G$ we notice that
by \eqref{DD.7a}
   \begin{equation}
    \label{H.1.3}
   \int_{-\eps^{-1}}^{\eps^{-1}} u^*|R_\eps|    \le c\eps e^{-a _{\eqref{DD.4}}\eps^{-1}(1-x_0)}
    \end{equation}
a property which will be used in the sequel. We have already proved with \eqref{H.1} the first condition
for $h_0\in \mathcal G$.  Then, since $\dis{  \frac{d(h_{0} -h_{\eps})}{dx} = -\frac{dR_\eps}{dx}}$
it will suffice to show that
      \begin{equation}
   \label{e4.6.2.1}
|\frac{dR_\eps}{dx} |  \le c \eps^2 \text{\bf 1}_{x\ge \eps^{-1}-1}
    \end{equation}
We have
     \begin{equation}
   \label{eG.8}
\frac{dR_\eps}{dx}  =
 \int_{\eps^{-1}}^{\eps^{-1}+1}
J(x,y)[\psi(y)-\psi(2\eps^{-1}-y)]\,dy,\quad \psi = \frac{dm^*}{dx}
  \end{equation}

To bound the term $|\psi(x)-\psi(x+ \xi)|$, $\xi=x'-x$, $x$ and $x'$ in  $[\eps^{-1}-1,\eps^{-1}+1]$,
in \eqref{eG.8} we use the expression \eqref{H.00} for $\psi$.  By \eqref{DD.28}
   \begin{equation}
    \label{H.01}
\|L^{-1} \phi + \sum_{n=0}^N (A^*)^n\phi\| \le  c' \|\phi\| e^{-a_{\eqref{DD.28}} N} \le c'' \eps^3,\quad
\phi= (\eps j)-(\eps j \int u^* ) u^*
    \end{equation}
if $N= C \log \eps^{-1}$ with $C$ large enough.  We have
$(A^*)^n\phi=(A^*)^n(\eps j)- (\eps j \int u^* ) (A^*)^nu^*$
By \eqref{DD.7a} and since $\la^*\in (0,1)$
   \begin{equation*}
(A^*)^n u^*(x) \le  u^*(x) \le c_{\eqref{DD.7a}} e^{-a _{\eqref{DD.4}}|x-\eps^{-1}x_0|}
    \end{equation*}
so that $|\sum_{n=0}^N\{(A^*)^n\phi - (A^*)^n(\eps j)\}| \le c\eps^3$
for $x\ge \eps^{-1}-1$ and $\eps$ small enough.  \eqref{e4.6.2.1} will then follow
from
    \begin{equation}
    \label{eG.133}
\sum_{n=0}^N|\int (A^*)^n (x,y)(\eps j)dy-\int (A^*)^n(x',y)(\eps j)dy|\le c\eps^2
    \end{equation}
 $x$ and $x'$ in  $[\eps^{-1}-1,\eps^{-1}+1]$.
To prove \eqref{eG.133} we write $\xi=x'-x$ and
    \begin{eqnarray*}
|\int  (A^*)^n (x,y)-\int (A^*)^n(x',y)| &\le&  \int A^*(x,x_1)\cdots A^*(x_{n-1},x_n) \big|1-\prod \frac {p^*(x_i +\xi)}{p^*(x_i)}\big|\\&\le& c n \eps b^n,\;\; b<1
    \end{eqnarray*}
as all points above are in $\{x: x -\eps^{-1} x_0
> \eps^{-1} - (N+1)\}$ (as $n\le N$) and in such a region
$0<c_{\eqref{e4.0}}<p^* <b<1$  (as $m^*>m_\beta)$ and  $|p^*(x_i +\xi)-p^*(x_i)|\le c\eps$, by Lemma \ref{lemma4.1}.  \eqref{eG.133}
is thus proved.
 \qed

\vskip1cm

  \begin{prop}
  \label{propH.2}
There are $c_{\eqref{e4.6.2.2.1}}$ and  $c_{\eqref{e4.6.2.2.1b}}$ so that for all
$\eps$ small enough the following holds.
Suppose that for $n\ge 1$, both $h_n$ and $h_{n-1}$ are in $\mathcal G$, then
      \begin{equation}
   \label{e4.6.2.2.1}
 N\big(m_n-m_{n-1}\big) \le c_{\eqref{e4.6.2.2.1}}  N\big(h_n-h_{n-1}\big)
    \end{equation}
where $m_i= \tanh\{\beta J^{\rm neum}*m_i+\beta h_i\}$, $i=n-1,n$.
Moreover
      \begin{equation}
   \label{e4.6.2.2.1b}
 N\big(m_0-m_{\eps}\big) \le c_{\eqref{e4.6.2.2.1b}}  \eps
    \end{equation}

  \end{prop}
\vskip.5cm

\noindent {\bf Proof.}  We first prove \eqref{e4.6.2.2.1} where we recall that $n\ge 1$.
Let $t\in [0,1]$ and
$h(t)=th_n+(1-t)h_{n-1}$.  Since $h_n$ and $h_{n-1}$ are
in $\mathcal G$ then, by convexity, $h(t)\in \mathcal G$ and
by Proposition \ref{propF.1}  there is
$m(t)$ such that $(h(t),m(t))\in \mathcal  A$, in particular
$m(t)=\tanh\{\beta J^{\rm neum}*m(t)+\beta h(t)\}$ and $m(0)=m_{n-1}$,
$m(1)=m_n$ so that
$\dis{|m_n-m_{n-1}|\le \sup_{t\in [0,1]}
|\frac{dm(t)}{dt}|}$. By  \eqref{E.3},  \eqref{DD.7a} and \eqref{DD.29}, recalling that $p_t\le \beta$ and writing
          \begin{equation}
    \label{H.5.1}
  \psi(x) :=  \Big(|\int u(t)[h_n-h_{n-1}]|\Big)\;
  e^{-a _{\eqref{DD.4}}|x-\eps^{-1}x_0|}
    \end{equation}
    \begin{equation}
    \label{H.6.1}
  |\frac{dm(t)}{dt}|  \le c\eps^{-1} \psi +c \int e^{- a_{\eqref{DD.29}}|x-y|}
\big(|h_n-h_{n-1}|(y) + \psi(y)\big)\, dy
         \end{equation}
We are going to prove that
           \begin{equation}
                \label{HHH}
 \psi \le c_{\eqref{HHH}} \eps^{10} e^{-a _{\eqref{DD.4}}|x-\eps^{-1}x_0|}N(h_n-h_{n-1})
         \end{equation}
By the definition of $\mathcal G$, $\dis{\int  u^*   h_i  =0}$, $i=n-1, n$, then
            \begin{equation}
                \label{H.6.1.1.0}
 \int u(t)[h_n-h_{n-1}]=
\int [u(t)-u^*]  [h_n-h_{n-1}]
         \end{equation}
       \begin{equation*}
       | \int [u(t)-u^*]  [h_n-h_{n-1}]|\le  N(h_n-h_{n-1})\int |u(t)-u^*| E_\eps^{-1}
       \le c\eps^{10}  N(h_n-h_{n-1})
           \end{equation*}
(by Corollary \ref{coroF.1}).  \eqref{HHH} is proved.
Using  \eqref{HHH} we have
       \begin{equation*}
\int e^{- a_{\eqref{DD.29}}|x-y|}
  \psi(y)\, dy \le c \eps^{10}
  e^{-a|x-\eps^{-1}x_0|}N(h_n-h_{n-1}),\quad a=\min\{a_{\eqref{DD.4}},a_{\eqref{DD.29}}\}
       \end{equation*}
       The other integral  on  the r.h.s.\ of \eqref{H.6.1} is bounded by
              \begin{eqnarray*}
\int e^{- a_{\eqref{DD.29}}|x-y|}
|h_n-h_{n-1}|(y)\, dy &\le&   N(h_n-h_{n-1})\int e^{- a_{\eqref{DD.29}}|x-y|} E_\eps(y)\,dy
  \\&\le& c e^{-a_{\eqref{e4.8a}}|x-\eps^{-1}x_0|}
       \end{eqnarray*}
because  $a_{\eqref{DD.29}}>
a_{\eqref{e4.8a}}$.  Collecting all these bounds we have from \eqref{H.6.1}
   \begin{equation*}
  |\frac{dm(t)}{dt}|(x) \le c\Big( \eps^{9}
  e^{-a_{\eqref{e4.8a}}|x-\eps^{-1}x_0|} + (1+ \eps^{10} ) e^{-a_{\eqref{e4.8a}}|x-\eps^{-1}x_0|}\Big)
   N(h_n-h_{n-1})
         \end{equation*}
which proves \eqref{e4.6.2.2.1}.

\noindent The proof
of \eqref{e4.6.2.2.1b} goes in the same way except
for \eqref{H.6.1.1.0} which becomes
            \begin{equation}
                \label{H.6.1.1.0.1}
 \int u(t)[h_0-h_{\eps}]=
\int [u(t)-u^*]  [h_0-h_{\eps}] +\int u^*h_{\eps}
         \end{equation}
By \eqref{H.1.1} and \eqref{H.1.3}  the latter integral is bounded by $
\le ce^{-a _{\eqref{DD.4}}\eps^{-1}(1-x_0)}$ and the bound \eqref{e4.6.2.2.1b}
is not affected.

 \qed

\vskip.5cm

  \begin{prop}
  \label{propH.4}
There is  $c_{\eqref{H.4}}\ge c_{\eqref{H.1}}$ so that for all $\eps$ small enough
the following holds. Given any $n\ge 0$ if
$h_k$, $k\le n$, is well defined and in $\mathcal G$ then also $h_{n+1}$ is well
defined and
   \begin{equation}
    \label{H.4}
    N(h_{k+1}-h_k) \le \begin{cases}c_{\eqref{H.4}} \eps N(h_k-h_{k-1}), & k=1,..,n\\
    c_{\eqref{H.4}} \eps, & k=0
    \end{cases}
    \end{equation}

  \end{prop}

\vskip.5cm

\noindent {\bf Proof.} By Proposition \ref{propF.1} there is  $m_k$,  $0\le k\le n$, so that
$(h_k,m_k)\in \mathcal A$ and, by \eqref{DD.5z},
$p_k\equiv p_{h_k,m_k} \ge C_{\eqref{DD.5z}}$.  As a consequence $p_n^{-1}$ is bounded and $h_{n+1}$
is well defined; moreover
$|p^{-1}_{k}-p^{-1}_{k-1}|
\le c |m_k-m_{k-1}|$ and  (for $x>\eps^{-1}x_0$)
   \begin{equation*}
   |h_{k+1}-h_k|  \le c \eps\Big( f + \frac{ \int  u^*   f}{ \int  u^*} \Big),\quad
   f(x)=\int_{\eps^{-1}x_0}^x |m_k-m_{k-1}| \, dy
         \end{equation*}
Let $x>\eps^{-1}x_0$, then by \eqref{e4.6.2.2.1} for $k\ge 1$
      \begin{equation}
   \label{H.5}
 e^{a_{\eqref{e4.8a}} (\eps^{-1}- x)}  f(x)= \{\int_{\eps^{-1}x_0}^x e^{-a_{\eqref{e4.8a}} (x - y)}
  c_{\eqref{e4.6.2.2.1}} \} N\big(h_k-h_{k-1}\big) \le c N\big(h_k-h_{k-1}\big)
    \end{equation}
and by \eqref{DD.7a}
      \begin{eqnarray}
    \nn
\int_{ \eps^{-1}x_0}^{\eps^{-1}} u^*f &\le&  c N\big(h_k-h_{k-1}\big) \int_{ \eps^{-1}x_0}^{\eps^{-1}}
  e^{-a_{\eqref{e4.8a}} (\eps^{-1}- x)}  e^{-a _{\eqref{DD.4}}|x-\eps^{-1}x_0|} \\&\le&  c N\big(h_k-h_{k-1}\big)
   e^{-a_{\eqref{e4.8a}} \eps^{-1}(1- x_0)}
     \label{H.5.0}
    \end{eqnarray}
By  \eqref{H.1.111} $\int u^*$ is bounded away from 0 hence the bound in \eqref{H.4}
for $k>0$ and  $x\ge \eps^{-1}x_0$.
When $k=0$ we use \eqref{e4.6.2.2.1b} after bounding $|m_0-m_\eps| \le N(m_0-m_\eps) E_\eps^{-1}$.
Analogous bounds hold for $x<\eps^{-1}x_0$ and \eqref{H.4} is proved.  \qed

\vskip.5cm

  \begin{prop}
  \label{propH.5}
In the same context of Proposition \ref{propH.4}, for any $k\le n+1$
   \begin{equation}
    \label{H.7}
    N(m_k-m_\eps)\le c
 \eps,\qquad
N(h_k-h_\eps)\le c'
 \eps
    \end{equation}
where $\dis{c= c_{\eqref{e4.6.2.2.1b}} + \frac{c_{\eqref{e4.6.2.2.1}}c_{\eqref{H.4}}}
    {1-\eps c_{\eqref{H.4}} }}$,  $\dis{c'= c_{\eqref{H.4}}(1+ \frac{1}{1-\eps c_{\eqref{H.4}} })}$. Moreover
       \begin{equation}
    \label{H.8}
N(\frac{d(h_k-h_\eps)}{dx}) \le c \eps^2
    \end{equation}
and, in particular,  $h_{n+1}\in \mathcal G$.

  \end{prop}

\vskip.5cm

\noindent {\bf Proof.} By  \eqref{H.4} for $i\ge 0$,  $N(h_{i+1}-h_{i}) \le
(\eps c_{\eqref{H.4}})^{i+1}$ and by \eqref{H.1},  $N(h_{0}-h_{-1}) \le \eps
c_{\eqref{H.1}}\le \eps c_{\eqref{H.4}}$,
 $h_{-1}=h_\eps$.  Then
   \begin{equation*}
N(h_k-h_\eps)\le \sum_{i=0}^k N(h_i-h_{i-1}) \le \eps c_{\eqref{H.4}}(1+ \frac{1}{1-\eps c_{\eqref{H.4}} })
    \end{equation*}
hence
the statement in \eqref{H.7} about $h_k$.  The one about $m_k$ is proved similarly, using
\eqref{e4.6.2.2.1b} and \eqref{e4.6.2.2.1}.
To prove \eqref{H.8} we write
       \begin{equation}
    \label{H.9}
 |\frac{d(h_k-h_{k-1})}{dx}| \le c \eps |m_{k-1}-m_{k-2}| \le c' \eps |h_{k-1}-h_{k-2}|
     \end{equation}
  so that by \eqref{e4.6.2.1}, \eqref{H.4} and \eqref{H.1} and with $h_{-1}:=h_\eps$, for $x>\eps^{-1}x_0$,
         \begin{equation*}
e^{a_{\eqref{e4.8a}} (\eps^{-1}- x)} |\frac{d(h_k-h_{\eps})}{dx}| \le e^{a_{\eqref{e4.8a}} (\eps^{-1}- x)} |\frac{d(h_0-h_{\eps})}{dx}|+
c'\eps \sum_{i=1}^{k-1}N\big( |h_{i}-h_{i-1}|\big) \le c''\eps^2
     \end{equation*}
An analogous bound holds for  $x<\eps^{-1}x_0$ hence \eqref{H.8}.  \qed

\vskip1cm

{\bf Conclusion of the  proof of Theorem \ref{thme2.2}.}  We shall first prove by induction that  $h_n\in \mathcal G$
for all $n$. Indeed  $h_0\in \mathcal G$
by Proposition \ref{propH.1} and by
Proposition \ref{propH.4} if $h_k\in  \mathcal G$ for all $k\le n$, then
$h_{n+1}\in \mathcal G$.  Thus  $h_n\in \mathcal G$
for all $n$ and by Proposition \ref{propF.1}
there is  $m_n$ so that  $(h_n,m_n)\in \mathcal A$.  We shall next prove
that $(h_n,m_n)$ converges
in sup norm to a limit $(h,m)$ and that,
writing $h_{-1}=h_\eps$ and $m_{-1}=m_\eps$,
  \begin{equation*}
h= h_\eps+ \sum_{n=0}^\infty (h_n-h_{n-1}),\qquad m= m_\eps+ \sum_{n=0}^\infty (m_n-m_{n-1})
    \end{equation*}
The first series in fact
converges because
$N(h_{n+1}-h_{n}) \le (c_{\eqref{H.4}}\eps)^{n+1}$,
as remarked in the proof of Proposition \ref{propH.5}.  The series for $m$ converges
for the same reason because
$N(m_n-m_{n-1}) \le c_{\eqref{e4.6.2.2.1}}N(h_n-h_{n-1})$.
By \eqref{H.7}, $N(m-m_\eps) \le c\eps$ and  $N(h-h_\eps) \le c\eps$; moreover
  \begin{eqnarray*}
&& m=\lim_{n\to \infty}m_n = \lim_{n\to \infty}\tanh\{\beta J^{\rm neum}*m_n+\beta h_n\}=
\tanh\{\beta J^{\rm neum}*m+\beta h\}
\\
&& h=\hat h - \frac{\int \hat h u^*}{\int u^*},\quad \hat h(x)=\int_{\eps^{-1}x_0}^x \frac{-\eps j}{\chi(m)}
    \end{eqnarray*}
because $\dis{h=\lim_{n\to \infty}\{\hat h_n - \frac{\int \hat h_n u^*}{\int u^*}\},\quad \hat h_n(x)=\int_{\eps^{-1}x_0}^x \frac{-\eps j}{\chi(m_n)}}$.  As a consequence, for any $z\in \eps^{-1}(-1,1)$,
      \begin{equation*}
h(x) = h(z)+ \int_z^x \frac{-\eps j}{\chi(m)}
    \end{equation*}
so that the proof of Theorem  \ref{thme2.2} will be complete once we  show that:

\vskip.3cm

\centerline{$\bullet$\; there is $x_\eps$ such that $h(x_\eps)=0$\qquad
$\bullet$\,  $\dis{\lim_{\eps\to 0}\eps x_\eps= x_0}$}

\vskip.3cm
\noindent The existence of $x_\eps$ is proved using
the implicit function theorem. We thus want to prove that $h(\eps^{-1}x_0)$
is ``small''. Since
$\hat h (\eps^{-1}x_0)=0$   we need to control
$|\int \hat h u^*|$. We write $\int \hat h u^*= \int(\hat h-\hat h_n)u^*+\int(\hat h_n-\hat h_0)u^*
+\int \hat h_0u^*$.  The first term vanishes as $n\to \infty$
because $\int u^*<\infty$ and for any $x>\eps^{-1}x_0$ (for instance)
  \begin{eqnarray*}
|\hat h(x)-\hat h_n(x)|&\le& |\eps j|\int_{\eps^{-1}x_0}^x |\chi(m)^{-1}-\chi(m_{n-1})^{-1}|
\le c \eps |x-\eps^{-1}x_0| \|m-m_{n-1}\| \\&\le & c'\|m-m_{n-1}\| \to 0
   \end{eqnarray*}
having used that  $\chi(m_n)=p_{h_n,m_n} \ge C_{\eqref{DD.5z}}$ and therefore
$\chi(m)  \ge C_{\eqref{DD.5z}}$ as
$m_n\to m$ in sup norm. Analogously
  $|\chi(m_{n-1})^{-1}-\chi(m_{\eps})^{-1}| \le c|m_{n-1}-m_{\eps}| \le c N(m_{n-1}-m_{\eps}) E_\eps^{-1}\le c'\eps E_\eps^{-1}$, so that
for $x>\eps^{-1}x_0$
   \begin{eqnarray*}
|\hat h_n(x)-\hat h_0(x)|&\le&|\eps j|\int_{\eps^{-1}x_0}^x |\chi(m_{n-1})^{-1}-\chi(m_\eps)^{-1}|
\le c \eps^2 \int_{\eps^{-1}x_0}^x E_\eps(y)^{-1} \\&\le &  c  \eps^2|x-\eps^{-1}x_0|e^{-a_{\eqref{e4.8a}}\eps^{-1}(1-x)}
   \end{eqnarray*}
Thus by \eqref{DD.7a} the second term is bounded by $\int|\hat h_n-\hat h_0|u^*\le
c \eps^2 e^{-a_{\eqref{e4.8a}} (1-x_0)\eps^{-1}}$.
Finally, since  by \eqref{eG.1} $\hat h_{0} = h^*$,  by \eqref{H.1.1}
$\dis{   \int_{-\eps^{-1}}^{\eps^{-1}}\hat h_{0} u^*
    \le c_0e^{-a _{\eqref{DD.4}}\eps^{-1}(1-x_0)}}$.
In conclusion, letting $n\to \infty$,
     \begin{equation*}
|\int u^* \hat h | \le c\eps^2  e^{-a_{\eqref{e4.8a}} (1-x_0)\eps^{-1}},\quad
|h(\eps^{-1}x_0) | \le c'\eps^2  e^{-a_{\eqref{e4.8a}} (1-x_0)\eps^{-1}}
     \end{equation*}
We shall next prove that $h$ is continuous and that it changes
sign in a small interval around $\eps^{-1}x_0$, thus concluding
that there is $x_\eps$ in such interval where  $h$
vanishes. We have
$\dis{\frac{dh}{dx}(x)= \frac{-\eps j}{\beta(1-m^2(x))}}$
which, by \eqref{DD.5z}, is bounded. Moreover
$\dis{N(m-m_\eps) \le c
\eps}$ and $|m_\eps-\bar m_{x_0}|\le c\eps$ in $[\eps^{-1}x_0-1,\eps^{-1}x_0+1]$.
In such interval therefore $\dis{|\frac{dh}{dx}(x) |\ge a\eps}$,
$a>0$. Hence there is $x_\eps$ where $h(x_\eps)=0$ and
       \begin{equation}
    \label{eG.26}
 |x_\eps-\eps^{-1}x_0| \le c''\eps e^{-a_{\eqref{e4.8a}} (1-x_0)\eps^{-1}}
     \end{equation}

\vskip1cm

\noindent {\bf Acknowledgments.} This work was started  during a
workshop in Athens, Nov. 2008, from discussions with N. Alikakos and
G. Fusco, to whom we are deeply indebted.  We also acknowledge very
kind hospitality at the Mathematical Department of Athens
University.  
The research of D. T. was partially supported by the Marie-Curie grant PIEF-GA-2008-220385. 
We finally thank Enza Orlandi for very helpful discussions
and for showing us her unpublished notes on the dependence
of critical points of the L-P functional
on the boundary conditions.

\vskip1cm

\bibliographystyle{amsalpha}

 \end{document}